\documentclass[aps,prb,twocolumn,superscriptaddress,floatfix,longbibliography]{revtex4-2}

\usepackage{graphicx,graphics}
\usepackage{dcolumn}
\usepackage{amsmath}
\usepackage{amsfonts}
\usepackage{amssymb}
\usepackage{latexsym,verbatim}
\usepackage{physics}
\usepackage{bm}
\usepackage{color}
\usepackage{xcolor}
\usepackage[normalem]{ulem}
\usepackage[percent]{overpic}
\usepackage{bbold}
\usepackage{lipsum}
\usepackage{ragged2e}
\usepackage[breaklinks=true,colorlinks,citecolor=blue,linkcolor=blue,urlcolor=blue]{hyperref}

\usepackage{graphicx}
\graphicspath{{figures_old}}

\usepackage{nomencl}
\usepackage{ulem}
\makenomenclature

\DeclareMathAlphabet\mathbfcal{OMS}{cmsy}{b}{n}

\usepackage{soul}
\usepackage{orcidlink}

\begin{document}

\title{Interplay between photon condensation and electron-electron interactions in molecular systems}

\author{Matteo Parisi \orcidlink{0009-0001-7530-7344}}
\affiliation{Dipartimento di Fisica e Astronomia ``Ettore Majorana'', Universit\`a di Catania, Via S. Sofia 64, I-95123 Catania,~Italy}
\author{Elisabetta Paladino \orcidlink{0000-0002-9929-3768}}%
\affiliation{Dipartimento di Fisica e Astronomia ``Ettore Majorana'', Universit\`a di Catania, Via S. Sofia 64, I-95123 Catania,~Italy}
\affiliation{INFN, Sez.~Catania, I-95123 Catania,~Italy}
\author{Giuseppe A. Falci \orcidlink{0000-0001-5842-2677}}%
\affiliation{Dipartimento di Fisica e Astronomia ``Ettore Majorana'', Universit\`a di Catania, Via S. Sofia 64, I-95123 Catania,~Italy}
\affiliation{INFN, Sez.~Catania, I-95123 Catania,~Italy}
\author{Gian Marcello Andolina \orcidlink{0000-0002-4219-7177}}
\affiliation{JEIP, UAR 3573 CNRS, College de France, PSL Research University, 11 Place Marcelin Berthelot, F-75321 Paris, France}
\author{Salvatore Savasta \orcidlink{0000-0002-9253-3597}}
\affiliation{Dipartimento di Scienze Matematiche e Informatiche, Scienze Fisiche e Scienze della Terra, Università di Messina, I-98166 Messina, Italy
}
\author{Marco Polini~\orcidlink{0000-0003-2217-7408}}
\affiliation{Dipartimento di Fisica dell’Universit\`a di Pisa, Largo Bruno Pontecorvo 3, I-56127 Pisa, Italy}
\affiliation{ICFO-Institut de Ci\`{e}ncies Fot\`{o}niques, The Barcelona Institute of Science and Technology, Av. Carl Friedrich Gauss 3, 08860 Castelldefels (Barcelona),~Spain}
\author{Francesco M. D.~Pellegrino~\orcidlink{0000-0001-5425-1292}}
\affiliation{Dipartimento di Fisica e Astronomia ``Ettore Majorana'', Universit\`a di Catania, Via S. Sofia 64, I-95123 Catania,~Italy}
\affiliation{INFN, Sez.~Catania, I-95123 Catania,~Italy}
%

\begin{abstract}
We investigate a minimal molecular model consisting of square planar plaquettes hosting multiple electrons, whose dynamics is governed by a tight-binding Hamiltonian supplemented by on-site Hubbard repulsion. By coupling this system to a spatially nonuniform cavity mode, we analyze the emergence of a magnetostatic instability, namely photon condensation, originating from the paramagnetic Van Vleck mechanism.
The global behavior of the system is analyzed for different electronic filling factors, and we find that, except for the special cases of half-filling and a single electron, where the transition, if it occurs, is necessarily a second-order phase transition, the global system may also undergo a first-order transition because of the action of the electron-electron interaction.
The polaritonic excitation energies are analyzed, providing clear spectroscopic signatures of the magnetostatic instability and of its order.
\end{abstract}

\maketitle

\section{Introduction}

A central goal of the cavity-control paradigm is to engineer the properties of quantum materials by embedding them in cavities~\cite{schlawin_apr_2022,mivehvar_advphysics_2021,Bloch_Nature_2022,GarciaVidal_science_2021}. In the strong-coupling regime~\cite{kockum_natrevphys_2019}, vacuum fluctuations can significantly modify the material properties.
It has been seen experimentally how cavity embedding changes the magnetotransport properties in two-dimensional (2D) systems~\cite{Paravicini-Bagliani_natphys_2019}, breaks the topological protection in the integer quantum Hall effect~\cite{Appugliese_science_2022}, and alters the critical temperature of a charge density wave transition~\cite{jarc_nature_2023,chiriaco_prb_2024}.
In addition, further analysis has been devoted to studying how quantum fluctuations in the cavity field can be exploited to manipulate electronic instabilities and ordered phases, such as superconductivity~\cite{sentef_scienceadv_2018,curtis_prl_2019,schlawin_prl_2019,andolina_prb_2024} and ferroelectricity~\cite{latini_pnas_2021,ashida_prx_2020}, and to modify the electronic topological characteristics of materials~\cite{ciuti_prb_2021,chiocchetta_natcomm_2021,dmytruk_commphys_2022,bacciconi_prb_2024,winter_prb_2025}.
Moreover, cavity coupling has also been proposed as a route to induce genuine phase transitions involving both light and matter through the emergence of an equilibrium superradiant phase~\cite{Kirton_AQT_2018}, characterized by a macroscopically large number of coherent photons.
In the non-equilibrium regime, this phase transition has been observed in pumped ultracold gases trapped in an ultrahigh-finesse optical cavity~\cite{baumann_nature_2010}, whereas its equilibrium counterpart remains an open problem, motivating the ongoing search for an appropriate physical platform~\cite{Kirton_AQT_2018}.
To distinguish it from non-equilibrium superradiant emission~\cite{gross_physrep_1982}, we refer to the equilibrium superradiant phase transition as photon condensation~\cite{Andolina_APS_2019, Andolina_APS_2020,Andolina_EJ+_2022}.
In electronic systems, gauge invariance imposes stringent constraints on the occurrence of this instability.
In particular, photon condensation is forbidden when the electromagnetic vector potential $\bm{A}$ is spatially uniform~\cite{duncan_pra_1974,Rzaewski_PRL_1975,BialynickiBirula_APS_1979,gawedzki_pra_1981,Andolina_APS_2020,Nataf_PRL_2019}, whereas it can occur in the presence of a transverse, spatially nonuniform vector potential $\bm{A}(\bm{r})$ with $\nabla\times\bm{A}(\bm{r})\neq0$~\cite{Nataf_PRL_2019,Andolina_APS_2020,guerci_prl_2020,guerci_prb_2021,roman-roche_prl_2021,manzanares_prb_2022}.
In this case, the photon-condensed phase can emerge as a magnetostatic instability driven by the orbital paramagnetic response of the electronic system, closely related to the Condon magnetostatic instability~\cite{condon_prb_1966}.

Drawing inspiration from the intrinsically quantum-mechanical mechanism that gives rise to orbital paramagnetism in molecular systems, known as Van Vleck paramagnetism~\cite{Van_Vleck_1932,Grosso_2014}, Mercurio et al.~\cite{Mercurio_PRR_2024} have introduced a toy model consisting of an ensemble of identical, few-site coplanar molecules, described within the tight-binding approximation, each containing a single itinerant electron and coupled in a gauge-invariant fashion through Peierls substitution~\cite{li_prb_2020,dmytruk_prb_2021,luttinger_pr_1951,kohn_prb_1959,alexandrov_prl_1991,graf_prb_1995,boykin_ajp_2001,ibanes-azpiroz_pra_2014} to a spatially non-uniform quantized electromagnetic mode.
Here, Mercurio et al.~\cite{Mercurio_PRR_2024} demonstrated that, when the molecules are in the Van Vleck paramagnetic regime, a photon condensation phase transition can occur as a second-order quantum phase transition, and the critical coupling constant is determined by the value of the magnetic susceptibility.

In this work, we extend the single-electron analysis of Ref.~\cite{Mercurio_PRR_2024} to molecular systems hosting multiple spinful itinerant electrons, by explicitly incorporating electron-electron interactions through an on-site Hubbard repulsion, which can play a critical role in systems of extremely small size~\cite{Bruus_2004}.
Analogously to Refs.~\cite{bacciconi_scipost_2023,Mercurio_PRR_2024}, we introduce a spatially dependent mode function $\bm{A}(\bm{r})$ such that, in the region close to the matter system, the associated magnetic field $\bm{B}=\nabla\times \bm{A}(\bm{r})$ can be approximated as spatially uniform.

Using a mean-field decoupling between photonic and matter degrees of freedom~\cite{Wang-Hioe_PRA_1973,nataf_natcomm_2010,guerci_prl_2020,roman-roche_prl_2021,Mercurio_PRR_2024}, we examine how electronic correlations affect the emergence of a photon condensation phase transition.
In contrast to the treatment of light-matter coupling, electron-electron repulsion is treated exactly through an exact diagonalization of the Hubbard Hamiltonian~\cite{Altland_book}, which governs the dynamics of electrons within the single plaquettes.
Depending on the chosen filling factor, the electron-electron interaction can either renormalize the critical coupling at which photon condensation emerges as a second-order phase transition, or alter the character of the transition so that it becomes a first-order phase transition.
The emergence of a photon-condensation phase transition in molecular systems is particularly intriguing in view of recent progress in coupling molecular matter to quantized electromagnetic fields~\cite{Koch_PRA_2006,Koch_Nature_2016}. Moreover, this possibility is especially relevant in the context of polaritonic chemistry~\cite{Basov_nanoph_2020,Bloch_Nature_2022,Fregoni_ACS_2022,Ebbesen_ACS_2023,Gonzales_PRL_2016,Mandal_Chem_rev_2023,Hirai2023,Gu_ACR_2023,Xiang2024,Ribeiro_CS_2018}, which exploits strong light-matter coupling between molecules and confined electromagnetic field modes to enable new chemical reactivities.

This paper is structured as follows. Section~\ref{sec:model} introduces the four-site Hubbard plaquette model, while Sec.~\ref{sec:spectrum_nocavity} reviews the main features of its many-electron spectrum, building on established results for the isolated Hubbard plaquette, which provide the basis for the subsequent analysis of the cavity-coupled system. In Sec.~\ref{sec:light-matter}, the molecular system is coupled to the quantized electromagnetic field through the Peierls substitution. 
Section~\ref{sec:mf_theory} investigates photon condensation within a mean-field treatment of the light-matter coupling, focusing on the role of 
electron-electron interactions and electronic filling in determining the nature of the transition. In Sec.~\ref{sec:van_vleck}, this behavior is 
related to the magnetic response of the molecular system and to the Van Vleck paramagnetic mechanism. Section~\ref{sec:polaritonic_spectrum} analyzes the 
collective excitations arising in the coupled light-matter system, i.e., polaritons, across both first- and second-order phase transitions, for different fixed values of the electron-electron repulsion.
Finally, conclusions are drawn in Section~\ref{sec:conclusions}.

\section{Model}
\label{sec:model}

\begin{figure}[t]
    \centering
    \includegraphics[width=0.9\columnwidth]{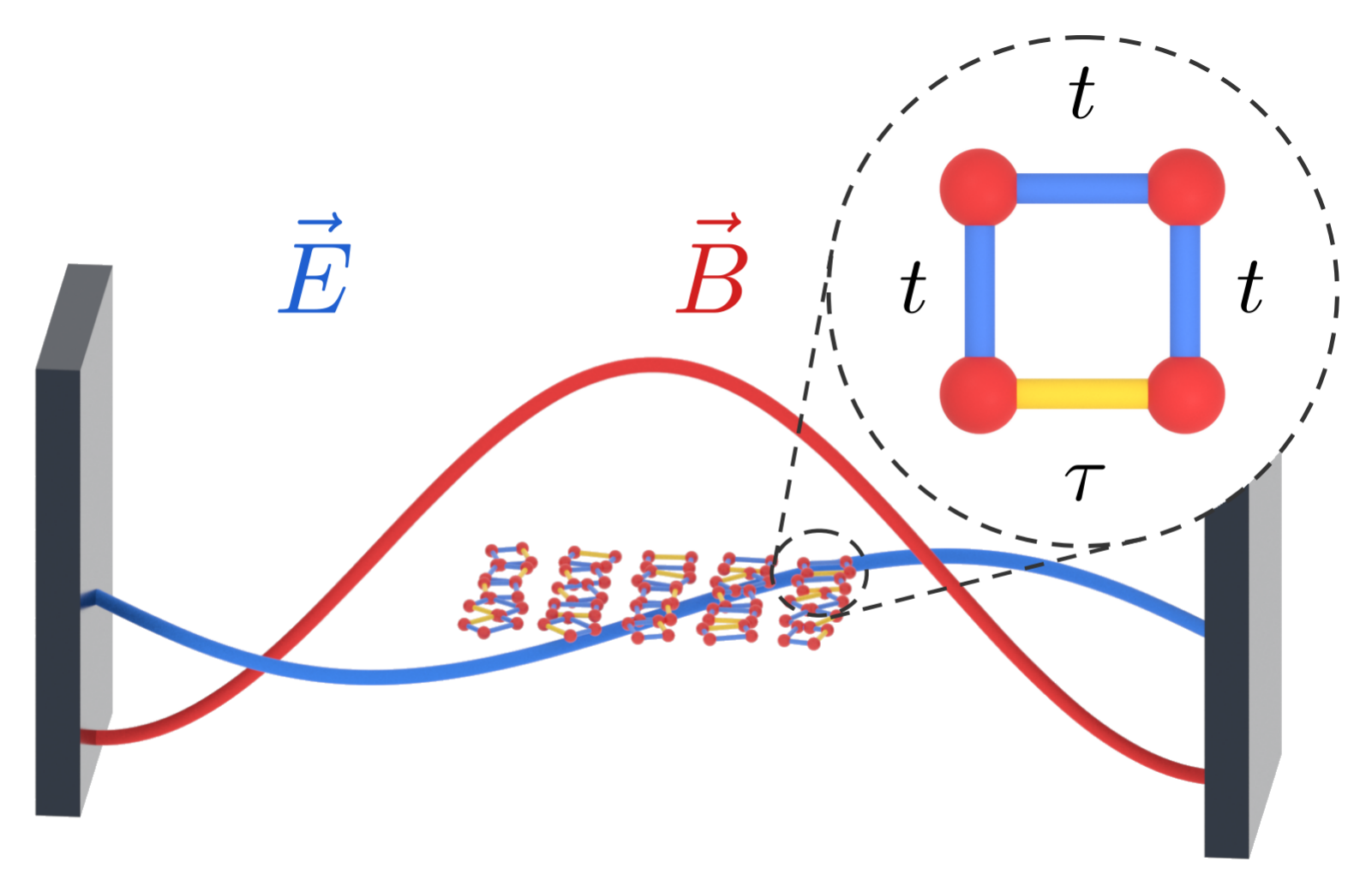}
    \caption{Schematic representation of the system considered. An ensemble of planar molecular plaquettes is placed in a cavity near a magnetic-field antinode and an electric-field node of the relevant cavity mode, such that the magnetic field is perpendicular to the molecular plane while the electric field lies within the plane. All plaquettes are assumed to lie in the $x$-$y$ plane. Arbitrary rotations about the $z$ axis do not affect their coupling to the perpendicular cavity magnetic field. The inset shows the four-site molecular plaquette, with hopping amplitude $t$ along three bonds and $\tau$ along the fourth bond.}
    \label{fig:cavity_scheme}
\end{figure}

We examine a toy  model to describe a system composed of $N$ identical planar molecules, each consisting of $n_{\rm s}$ atoms lying within the $x$-$y$ plane. Each molecule is modeled as a planar plaquette and described by a tight-binding Hamiltonian that includes only nearest-neighbor hopping processes and no spin-flip term. 
A schematic representation of the system and of the molecular plaquette is shown in Fig.~\ref{fig:cavity_scheme}. The molecules are placed near
a magnetic-field antinode and an electric-field node of the relevant cavity mode, so that the magnetic field is oriented perpendicular to the molecular plane, whereas the electric field remains within that plane.
The cavity mode and its coupling to the molecular system are discussed in detail in Sec.~\ref{sec:light-matter}.

The kinetic Hamiltonian of the system of $N$ molecules is then
\begin{subequations}\label{eq:kin}
\begin{align}
\hat{\cal T}_0&=\sum^N_{k=1}\hat{T}_{ k}~,\\
\hat{T}_{k} &= -\sum_{j = 0}^{n_{\rm s}-1} \sum_{s_z=\downarrow,\uparrow}(t_{j, j+1} \hat{c}_{j  s_z k}^{\dagger} \hat{c}_{j+1 s_z  k} + {\rm H.c.})~,
\end{align}
\end{subequations}
where the operator $\hat{c}_{j s_z  k}^{\dagger}$
($\hat{c}_{j  s_z k}$) creates (destroys) an electron with spin $z$-projection $s_z$ at the site $j$ of the $k$-th molecule. 
Following Ref.~\cite{Mercurio_PRR_2024}, we  impose periodic boundary conditions and assume that one of the tunneling amplitudes differs from the others, namely $t_{j, j+1}$ are given by $t_{j, j+1}=\delta_{j, 0} \tau +(1-\delta_{j, 0}) t$, where $\tau \in \mathbb{R}$, and we assume $t > 0$ without loss of generality. 
From here on out, we focus on the particular case of the square plaquette, $n_{\rm s}=4$. In this setting, the eigenenergies of the kinetic Hamiltonian of each molecule can be expressed in a simple closed form:
\begin{subequations}\label{eq:Tspectrum}
\begin{align}
	\epsilon_0&=\frac{-|t+\tau|-\sqrt{4 t^2+(t-\tau)^2}}{2}~,\\
	\epsilon_1&=\frac{-|t+\tau|+\sqrt{4 t^2+(t-\tau)^2}}{2}{\rm sgn}(\tau-t)~,\\
	\epsilon_2&=-\epsilon_1~,\\
	\epsilon_3&=-\epsilon_0~.
\end{align}
\end{subequations}
We observe that for $\tau=-t$, the spectrum exhibits degeneracies at  $\epsilon_0=\epsilon_1=-\sqrt{2}t$ and $\epsilon_2=\epsilon_3=\sqrt{2}t$. In contrast, for $\tau=t$, a degeneracy appears at $\epsilon_1=\epsilon_2=0$.
When each molecule hosts a single electron, the energy gap between the ground state and the first excited state is given by $|t+\tau|$ and vanishes in the limit $\tau \to -t$. The narrowing of the energy gap gives rise to an enhanced paramagnetic susceptibility via the Van Vleck mechanism \cite{Van_Vleck_1932}, thereby making this platform a candidate for sustaining photon condensation~\cite{Mercurio_PRR_2024}.
More specifically, Van Vleck paramagnetism originates from the 
virtual mixing, induced by an external magnetic field, between the molecular ground state and excited states. Within second-order perturbation 
theory, the resulting paramagnetic contribution to the ground-state 
energy is inversely proportional to the energy separation between 
the states coupled by the magnetic field. Consequently, when a 
magnetically active excited state approaches the ground state, the 
small energy separation strongly enhances the orbital paramagnetic 
susceptibility. In Appendix~\ref{app:van_vleck_paramagnetism} we briefly remind the reader about the microscopic theory of Van Vleck paramagnetism~\cite{Grosso_2014}.
When multiple electrons are introduced into the molecular system described above, their confinement in a reduced space can make the role of the electron-electron interactions significantly important~\cite{Heikkila2013}.
In this context, we model the Coulomb repulsion through an on-site Hubbard term~\cite{Altland_book}:
\begin{subequations}\label{eq:repulsion}
\begin{align}
\hat{\cal U} & = \sum^N_{k=1} \hat{U}_k~,\\
    \hat{U}_k &= U  \sum_{j= 0}^{n_{\rm s}-1} \hat{c}_{j  \uparrow k}^{\dagger} \hat{c}_{j  \uparrow k} \hat{c}_{j  \downarrow k}^{\dagger} \hat{c}_{j  \downarrow k}~,
    \end{align}
\end{subequations}
where $U>0$. The interaction term describes an on-site Coulomb repulsion between two electrons occupying the same lattice site and having opposite spin $z$-projection, in accordance with the Pauli exclusion principle. 
The Hubbard model is used here as a molecular prototype, since it provides a simple framework to capture the competition between electronic delocalization, governed by the kinetic term, and electron-electron interactions, without aiming at a quantitative description of a specific molecular system.

The total system Hamiltonian is written as 
\begin{equation}\label{eq:H}
	\hat{\cal H}_{\rm e}= \hat{\cal T}_0+ \hat{\cal U}=\sum^{N}_{k=1} \hat{H}_{k}~,
\end{equation}
where $\hat{H}_{k}=\hat{T}_k+\hat{U}_k$ represents the single-molecule Hamiltonian of the $k$-th molecule. To obtain the exact spectrum for a generic single-molecule Hamiltonian $\hat{H}_{k}$, one has to work in a many-body basis~\cite{Lin_AIP_1993, Honet_AIP_2022, Sharma_CPC_2015}, as the repulsion term represents a two-body operator.
In this case, due to the reduced size of the molecular structure, performing the exact diagonalization is feasible. Specifically, the dimension of the many-body basis for $N_{\rm e}$ spinful electrons distributed among $n_{\rm s}$ spatial orbitals is defined in terms of the binomial coefficient ${\cal N}=\binom{2 n_{\rm s}}{N_{\rm e}}$~\cite{Bruus_2004}.
For the generic $k$-th molecule, the eigenstates $|\Psi_{n k}\rangle$ solve the eigenvalues equation $\hat{H}_k |\Psi_{n k}\rangle = E_n|\Psi_{n k}\rangle$, where $E_0\le E_1\le \ldots \le E_{{\cal N}-1}$, $n = 0,..., {\cal N}-1$. 
Since the generic single-molecule Hamiltonian is SU$(2)$-invariant~\cite{Arovas_AR_2022, Tasaki_JoP_1998}, it is therefore convenient to determine its eigenstates by seeking a common eigenbasis of the Hamiltonian $\hat{H}_{k}$, the total spin operator $\hat{S}^2$, and the spin projection along the $z$ axis, $\hat{S}_{z}$.
Taking advantage of this symmetry in the system significantly reduces the computational cost of diagonalizing the molecular Hamiltonian.
%
Moreover, we observe that the interacting system described above can be fully understood by restricting the analysis to cases where the total number of electrons satisfies $N_{\rm e}\le n_{\rm s}$.
When the electron number $N_{\rm e}$ exceeds half-filling, $N_{\rm e}>n_{\rm s}$, one can use the particle-hole symmetry to map these situations onto cases with an electron number of $2n_{\rm s}-N_{\rm e}$.
In fact, performing the transformation $\hat c_{j s_z k} \to (-1)^j \hat d^\dagger_{j s_z k}$, 
%
\begin{subequations}
\begin{align}
\hat{T}_{k} &= -\sum_{j = 0}^{n_{\rm s}-1} \sum_{s_z=\downarrow,\uparrow}(t_{j, j+1} \hat{d}_{j  s_z k}^{\dagger} \hat{d}_{j+1 s_z  k} + {\rm H.c.})~,\\
    \hat U_k &= 
    U  \sum_{j= 0}^{n_{\rm s}-1} \hat{d}_{j  \uparrow k}^{\dagger} \hat{d}_{j  \uparrow k} \hat{d}_{j  \downarrow k}^{\dagger} \hat{d}_{j  \downarrow k}+U(N_{\rm e}-n_{\rm s})~,
\end{align}
\end{subequations}
The kinetic Hamiltonian remains formally unchanged, whereas the Hubbard interaction is modified by a constant energy shift which depends on the number of electrons per molecule $N_{\rm e}$. 
%
The case with a single electron in the molecule, i.e., $N_{\rm e}=1$, corresponds to a non-interacting system in which the eigenvalues of the kinetic Hamiltonian directly determine the spectrum. This case will not be discussed here, as it has already been analyzed in Ref.~\cite{Mercurio_PRR_2024}.

\section{Many-electron spectrum} 
\label{sec:spectrum_nocavity}
\begin{figure}[t]
	\centering
    \begin{overpic}[width=\columnwidth]{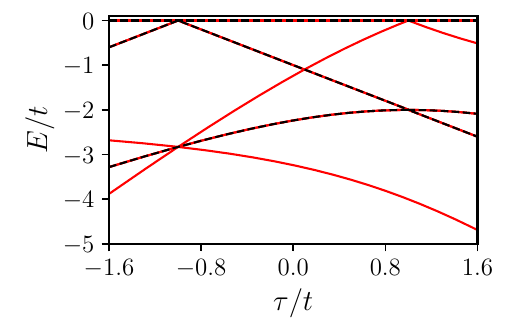}\put(2,52){a)}\end{overpic}\vspace{0em}
	    \begin{overpic}[width=\columnwidth]{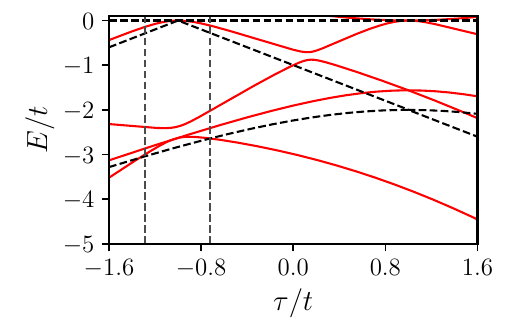}\put(2,52){b)}\end{overpic}\vspace{0em}  
	\caption{{
        a) Low energy spectrum for $N_{\rm e} = 2$ of a square molecule ($n_{\rm s}=4$), in the absence of repulsive electron-electron interaction. 
        b) Low-energy spectrum for the same molecular system, with $U/t = 1$. 
        The vertical dashed lines at $\tau_1\approx -1.33t$ and $\tau_2\approx -0.72t$ mark the boundaries where the ground state becomes a three-fold degenerate triplet state.
        In both panels, red solid lines represent the eigenenergies of the singlet states ($S=0$), while the black dashed lines refer to triplet states ($S=1$).}
	\label{fig:spec_U1_vs_U0}}
\end{figure}

\begin{figure}[t]
	\centering
    \begin{overpic}[width=0.9\columnwidth]{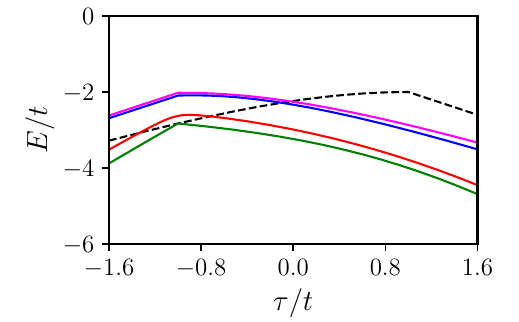}\put(0,52){a)}\end{overpic}\vspace{-0.em} 
    \begin{overpic}[width=0.9\columnwidth]{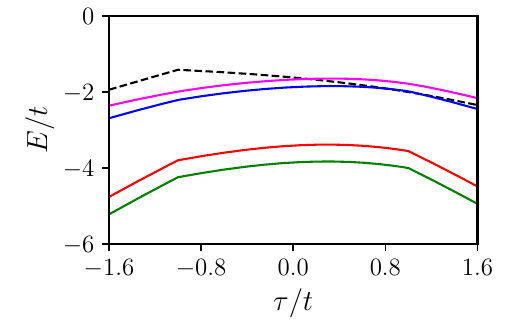}\put(0,52){b)}\end{overpic}\vspace{-0.em}  
	    \begin{overpic}[width=0.9\columnwidth]{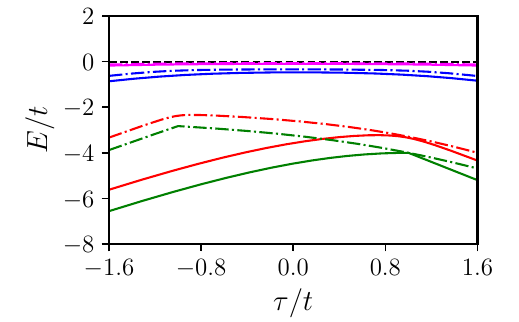}\put(0,52){c)}\end{overpic}\vspace{-0.5em} 
    \caption{The lowest eigenenergy, for a square molecular plaquette with $n_{\rm s}=4$, as a function of $\tau/t$, for $U = 0$ (red), $U/t = 1$ (green), $U/t = 20$ (blue), and $U/t = 100$ (magenta). 
    a) $N_{\rm e}=2$ electrons per molecule, the solid lines represent the lowest eigenenergy within the subspaces ${\bf{S}}$, and the minimal eigenenergy in the subspaces ${\bf{T}}_{i}$ ($i=0,\pm 1$) is $\epsilon_0+\epsilon_1$ (black dashed line), independently of $i$.
    b) $N_{\rm e}=3$ electrons per molecule, the solid lines indicate the lowest eigenenergy within the subspaces ${\bf{D}}_i$ ($i = \pm 1/2)$, while the minimum eigenenergy within the subspaces ${\bf{Q}}_i$ (where $i=\pm 1/2,\pm 3/2$) is denoted as $\epsilon_0$ (black dashed line), independently of $i$.
    c) Half-filling case, $N_{\rm e}=4$, colored lines refer to the singlet subspace ${\bf{S}}$ (solid lines) and triplet subspaces ${\bf{T}}_i$ (dash-dotted lines). Here, the zero-energy level (black dashed line) represents the eigenenergy of the five pentuplet states spanning the subspaces ${\bf{P}}_i$ ($i=0,\pm1,\pm 2$).
	\label{fig:spec_comp}}
\end{figure}

The properties of the isolated four-site Hubbard plaquette are well established in the literature~\cite{Arovas_AR_2022}. In this Section,  we review the features of its many-electron spectrum that will be relevant to the subsequent analysis of the cavity-coupled system.
Here, we employ the exact diagonalization approach to analyze the  spectrum of the interacting system governed by the Hamiltonian $\hat{H}_k$ of the generic $k$-th molecule with $n_{\rm s}=4$ sites.
We begin by examining the simplest case involving the electron-electron interaction, where each molecule contains $N_{\rm e}=2$ electrons.  Since $\hat{S}^2$ is a conserved quantity, the allowed total spins are $S = 0,1$, which correspond to singlet and triplet states, respectively, and the full Hilbert space of the molecule can be written as ${\bf {H}} = {\bf {T}} \oplus {\bf {S}}$. In particular, the possible spin projections within the triplet subspace ${\bf{T}}$ are $S_z =-1, 0, 1$, and the associated $S = 1$ subspaces with fixed $z$-spin projection $i$ are denoted with ${\bf{T}}_{i}$. Hence, the triplet subspace can be decomposed as a direct sum ${\bf {T}} = \bigoplus_{i=-1}^{1}{\bf{T}}_{i}$. The singlet subspace ${\bf {S}}$ is characterized by $S = 0$ and the only allowed $z$-spin projection is $S_z = 0$, therefore ${\bf {S}} = {\bf {S}}_0$.  Further details on the factorization of the Hilbert space can be found in App.~\ref{app:Hilbert_space_structure}.
For reference, we start by turning off the repulsive interaction. Fig.~\ref{fig:spec_U1_vs_U0}~a) shows the low-energy spectrum for a single square molecule with $N_{\rm e}=2$ electrons, in the absence of the interaction term $U$, as a function of the hopping parameter $\tau$. The red solid lines denote the spectrum associated with the singlet subspace. Black dashed lines indicate the eigenenergies associated with the triplet states, where each energy level is composed of three degenerate states with $z$-spin projections $S_z = 0, \pm 1$.
Specifically, we observe that when $\tau \neq -t$, the ground state is the $S_z = 0$ singlet $\hat{\varphi}^\dagger_{0 \uparrow k}\hat{\varphi}^\dagger_{0 \downarrow k} \ket{\emptyset_k}$ with energy $2\epsilon_0$. Here, $\ket{\emptyset_k}$ is the vacuum state for the $k$-th molecule, and the operator $\hat{\varphi}^\dagger_{\ell s_z k}$ creates an electron with a spin $z$-projection $s_z$ and occupying the $\ell$-th eigenstate of the kinetic energy operator $\hat T_k$. The first excitation energy is $\epsilon_0+\epsilon_1$, and it exhibits a four-fold degeneracy, comprising a $S_z = 0$ singlet, and the three degenerate components of the triplet with $S_z = 0, \pm 1$ spin projection along $z$. For $\tau\to -t$, $\epsilon_1$ converges to $\epsilon_0$, resulting in a ground state with six-fold degeneracy.
Regardless of the value of $\tau$, each degenerate triplet includes the states
\begin{subequations}\label{eq:triplets_2e}
\begin{align}
    \ket{T^{1}_{(\ell, m) k}} &= \hat{\varphi}^\dagger_{\ell \uparrow k} \hat{\varphi}^\dagger_{m \uparrow k}\ket{\emptyset_k}~,\\
    \ket{T^{0}_{(\ell, m) k}} &= \frac{1}{\sqrt{2}}(\hat{\varphi}^\dagger_{\ell \uparrow k} \hat{\varphi}^\dagger_{m \downarrow k} + \hat{\varphi}^\dagger_{\ell \downarrow k} \hat{\varphi}^\dagger_{m \uparrow k})\ket{\emptyset_k}~,\\
    \ket{T^{-1}_{(\ell, m) k}}  &= \hat{\varphi}^\dagger_{\ell \downarrow k} \hat{\varphi}^\dagger_{m \downarrow k}\ket{\emptyset_k}~,
\end{align}
\end{subequations}
with $\ell < m$. In particular, within the triplet subspace, the state with the lowest energy is obtained for $\ell = 0, m = 1$. As detailed in App.~\ref{app:Hilbert_space_structure}, within the triplet subspace, the orbital part of the two-electron wavefunction must be antisymmetric in order to respect the Pauli principle. The antisymmetric nature of the spatial component of the electronic wavefunction prevents the electrons from occupying the same lattice site, therefore, activating the Hubbard repulsion does not affect states belonging to the triplet subspace.
This can be observed by comparing the black dashed lines in Fig.~\ref{fig:spec_U1_vs_U0}~a) with those in Fig.~\ref{fig:spec_U1_vs_U0}~b), where we plot the low-energy spectrum for a molecule containing $N_{\rm e}=2$ electrons, with the repulsive interaction energy $U=t$. The dashed lines, denoting the triplet states, remain unchanged with respect to the $U = 0$ case, whereas the spectrum associated with the singlet subspace is modified by the repulsive interaction, because this configuration can support a double occupancy of one of the sites. 
Fig.~\ref{fig:spec_U1_vs_U0}~b) shows that there are two particular hopping parameters, $\tau_1<-t$ and $\tau_2>-t$, which depend on the repulsive energy $U$, such that for any value of $\tau$ where $\tau_1 \le \tau \le \tau_2$ the ground state of the system is the degenerate triplet of energy $\epsilon_0 + \epsilon_1$. In contrast, outside this specified range, the ground state is a non-degenerate singlet, 
resides in the spin sector $S_z=0$, and its energy value depends on $U$. 
This phenomenology can be better understood in Fig.~\ref{fig:spec_comp}~a), where the minimum eigenenergy in the singlet subspace $\bf S$ is illustrated as a function of $\tau/t$ for different values of $U\ge0$: $U = 0$ (red), $U/t = 1$ (green), $U/t = 20$ (blue), and $U/t = 100$ (magenta). The black dashed line indicates the lowest eigenenergy $\epsilon_0+\epsilon_1$ of the triplet subspace $\bf T$. This comparison indicates that for any value of $U$, there is a bounded range of $\tau$ within which the lowest eigenenergy is associated with $\bf T$, and the ground state exhibits a triple degeneracy, which includes the single $S_z = 0$ triplet $\ket{T^{0}_{(0, 1) k}}$, together with the two other degenerate, spin-polarized components of the triplet with spin projections $S_z=\pm 1$. We observe that the width of the interval $[\tau_1, \tau_2]$ increases as $U$ increases.

We now turn to an analogous analysis for the case where a single molecule contains $N_{\rm e}=3$ electrons.
In the three-electron regime, the total Hilbert space of the system can be written as a direct sum ${\bf H} = {\bf Q} \oplus {\bf D}$, where ${\bf Q}$ and ${\bf D}$ denote the quadruplet and doublet subspaces characterized by total spin $S = 3/2$ and $S = 1/2$, respectively.
Hence, these subspaces can be decomposed further as $\mathbf{Q} = \bigoplus_{i = -3/2}^{3/2} \mathbf{Q}_{i}$ and $\mathbf{D} = \bigoplus_{i = -1/2}^{1/2} \mathbf{D}_{i}$, where the index $i$ runs over all allowed values of the spin-projection quantum number $S_z$.
%
%
As in the case of the triplet for $N_{\rm e}=2$ electrons, the states belonging to the $S = 3/2$ subspace are unaffected by electron-electron repulsion, because their orbital wavefunction is antisymmetric (see App.~\ref{app:Hilbert_space_structure} for details).
Therefore, the states of the $S = 3/2$ sector are degenerate quadruplets having the four possible spin projections $S_z =\pm 3/2, \pm 1/2$, both in the non-interacting and interacting case. The fully polarized members of the quadruplet with energy $\epsilon_\ell+\epsilon_m+\epsilon_n$ can be written as 
\begin{equation}
    \ket{Q^{3s_z}_{(\ell, m, n) k}} = \hat{\varphi}^\dagger_{\ell s_z k} \hat{\varphi}^\dagger_{m s_z k}\hat{\varphi}^\dagger_{n s_z k}\ket{\emptyset_k}~,
\end{equation}
with $\ell<m<n$, and belong to the subspaces ${\bf Q}_{3s_z}$. 
For the fully spin-polarized sectors $S_z=\pm 3/2$, it is evident that the repulsive interaction is irrelevant, thanks to the Pauli principle. In turn, the unpolarized quadruplet states read
\begin{equation}\label{eq:quadruplets_3e}
\begin{aligned}
    \ket{Q^{s_z}_{(\ell, m, n) k}} &=
    \frac{1}{\sqrt{3}}(\hat{\varphi}^\dagger_{\ell s_z k}
    \hat{\varphi}^\dagger_{m s_z k}
    \hat{\varphi}^\dagger_{n -s_z k}+\\
    &+ \hat{\varphi}^\dagger_{\ell s_z k}
    \hat{\varphi}^\dagger_{m -s_z k}
    \hat{\varphi}^\dagger_{n s_z k}+\\
    &+ \hat{\varphi}^\dagger_{\ell -s_z k}
    \hat{\varphi}^\dagger_{m s_z k}
    \hat{\varphi}^\dagger_{n s_z k})
    \ket{\emptyset_k}~,
\end{aligned}
\end{equation}
and spans the subspaces ${\bf Q}_{\pm s_z}$.
%
%
Fig.~\ref{fig:spec_comp}~b) shows the lowest eigenenergy within the doublet subspace as a function of the inequivalent hopping $\tau/t$, for $U = 0$ (red), $U/t = 1$ (green), $U/t = 20$ (blue), and $U/t = 100$ (magenta). It also includes the eigenenergy $\epsilon_0$ (black dashed line), which corresponds to the minimum eigenenergy in the quadruplet subspace.
We observe that for intermediate values, $U< \bar{U}\sim 20 t$,  of the repulsive energy, the global minimum of the eigenenergy is linked to the doublet subspace ${\bf{D}}$, independently of $\tau$. In contrast, for significantly large values of electron-electron energy $U$, there is a restricted range of the hopping parameter $\tau$ in the vicinity of $\tau = t$, within which the ground state of the system is a degenerate quadruplet with energy $\epsilon_0$. We observe that, in this condition, two of the four degenerate ground states, namely the fully-polarized states $\ket{Q^{3s_z}_{(\ell, m, n) k}}$, display ferromagnetic order~\cite{Nagaoka_PR_1966, Tasaki_PRB_1989, Kollar_PRB_1996}. 
More precisely, a numerical verification indicates that the minimum value of $U$ necessary for crossing the lowest eigenenergies is $\bar{U}=18.6t$, in accordance with previous results~\cite{Arovas_AR_2022}. 
This analysis of the square molecular plaquette is concluded by evaluating the scenario of the half-filling, where $N_{\rm e}=4$. The possible total spins are $S = 0, 1, 2$, and the corresponding subspaces ${\bf {S}}, {\bf {T}}, {\bf {P}}$ are formed by singlets, triplets and pentuplets, respectively. Each subspace can be further decomposed in lower-dimensional sectors having a $z$-spin projections allowed for the specific value of $S$. Specifically, the singlet subspace ${\bf {S}}$ coincides with the ${\bf {S}}_0$ subspace of the singlet having $S_z = 0$ spin projection. Within the triplet subspace the possible spin projections along $z$ are $S_z = 0, \pm 1$ and it results $\bigoplus_{i = -1}^{1} {\bf {T}}_i$. Finally, the pentuplet subspace ${\bf{P}}$ includes states with spin projections $S_{z} = -2,-1,\dots,2$ and decomposes as $\bigoplus_{i = -2}^{2} {\bf {P}}_i$, where each ${\bf {P}}_i$ is one-dimensional. In particular, the fully polarized spin sectors ${\bf {P}}_{4s_z}$ consist of only a four-electron state, $\hat{\varphi}^\dagger_{0 s_z k}\hat{\varphi}^\dagger_{1 s_z k} \hat{\varphi}^\dagger_{2 s_z k}\hat{\varphi}^\dagger_{3 s_z k}\ket{\emptyset_k}$, and it is a member of the degenerate pentuplet with zero energy. The remaining three pentuplet states with $S_z = 0, \pm 1$ spin projection have eigenenergies that do not depend on the electron-electron repulsion $U$ as well, and their form is reported in App.~\ref{app:Hilbert_space_structure}. In turn, the eigenenergies of the other states, namely the singlet and triplet states, are influenced by $U$. 
These results are summarized in Fig.~\ref{fig:spec_comp}~c), where the minimum eigenenergies for the singlet (solid line) and triplet (dotted lines) are shown as a function of the hopping parameter $\tau/t$ for various values of $U$. The black dashed line marks the ground state of zero energy of the pentuplet subspace, which corresponds to the unique eigenenergy independent of the electron-electron interaction.
For every finite repulsive interaction $U$, it is noted that the minimal eigenenergy is associated with the singlet subspace ${\bf{S}}$ and this remains valid for $U=0$, except in the case where $t=\tau$ where the ground state exhibits a degeneracy, due to the presence of an eigenenergy of $2\epsilon_0 = -4t$ within the triplet subspace $\bf{T}$.

\section{Light-Matter interaction}
\label{sec:light-matter}
In this Section, we couple the electron system to a single quantized electromagnetic mode describing a uniform magnetic field along the $\hat{\bm{z}}$ direction, perpendicular to the molecular plane. In the symmetric gauge, the corresponding vector potential is $\bm{A}(\bm{r})=-B y \hat{\bm{x}} / 2+B x \hat{\bm{y}} / 2$, yielding $\bm{B}=\nabla \times \bm{A}(\bm{r})=B \hat{\bm{z}}$.
As an illustrative case, this configuration can be implemented in a rectangular cavity enclosed by perfectly conducting boundaries. In the Coulomb gauge, a cavity eigenmode exhibiting the required local field structure can be represented by the vector potential ${\bm A}(\bm r)=-\frac{B}{2k}\sin(k_z z)\big[\sin(k y)\bm u_x-\sin(k x)\bm u_y\big]$.
If the molecules are confined near $(x,y,z)=(0,0,\pi/(2k_z))$, the electric field vanishes at the localization center, while the magnetic field reaches an antinode with $\bm B=B\bm u_z$. Furthermore, the leading spatial variations of the magnetic field about this point arise only at second order in the displacement, so that the field may be treated as locally uniform over a sufficiently small molecular volume.
Throughout this work, we restrict the analysis to a single cavity mode,
retaining a low-energy mode with a spatial profile providing strong
magnetic coupling to the molecular system. This choice provides a
minimal description of the cavity sector, consistent with our main
purpose of isolating the effects of electron-electron interactions.

The interaction between the matter system and the cavity mode is implemented via Peierls minimal substitution (with $\hbar=1$)~\cite{dmytruk_prb_2021, Mercurio_PRR_2024}
\begin{equation}
	t_{j, j+1} \to t_{j, j+1} \exp\left( \frac{i e}{c}\int_{\bm{r}_j}^{\bm{r}_{j+1}} \bm{A}(\bm{r}) \cdot d \bm{r}\right)~.
\end{equation}
Quantization of the cavity field is obtained with the usual prescription which promotes $B$ from a {\it c}-number to a bosonic quantum operator  $B \to B_0 (\hat a +\hat a^{\dagger})$, where $\hat{a}^\dagger$ denotes the creation operator of a cavity photon and $\hat{a}$ represents its annihilation operator. 
The full Hamiltonian of the light-matter interacting system in a cavity assumes the form
\begin{subequations}
\begin{align}\label{eq:Hfull}
	\hat{\mathcal H}&=\omega_{\rm c} \hat a^\dagger \hat a+\hat{\cal T} + \hat{\cal U}~,\\
    	\hat{\cal T}&=-
        \sum_{k,j,s_z}(t_{j, j+1} e^{- i \eta(\hat a+\hat a^{\dagger})/\sqrt{N} } \hat{c}_{j  s_z k}^{\dagger} \hat{c}_{j+1 s_z  k}+{\rm H.c.})~,
\end{align}
\end{subequations}
where $\omega_{\rm c}$ is the cavity mode energy, $\eta=-2 \pi[\Phi /(\Phi_0 n_{\rm s})] \sqrt{N}$ quantifies the light-matter coupling strength and is proportional to $\Phi/\Phi_0 = B_0 A_{\rm p}/\Phi_0$, the ratio of the magnetic flux $\Phi$ to the quantum of flux $\Phi_0=2\pi  c/e$ that pierces the plaquette with area $A_{\rm{p}}$. 
The prefactor $\sqrt{N}$ is included in the definition of $\eta$ in order to make the thermodynamic scaling explicit. In fact, $B_0$ denotes the vacuum magnetic-field amplitude of the quantized cavity mode and, by electromagnetic-field normalization, it scales with the mode volume as $B_0\propto V^{-1/2}$. In the thermodynamic limit at fixed molecular density, one has $V\propto N$, which implies $B_0\propto N^{-1/2}$. Consequently, $\eta\propto B_0\sqrt{N}$ remains finite as $N\to\infty$.
Here, the Hubbard repulsive interaction acts as a local operator and remains unchanged by the introduction of the Peierls substitution~\cite{dmytruk_prb_2021, Andolina_EJ+_2022}.
%
%
\section{Mean-Field Theory of Photon Condensation}
\label{sec:mf_theory} 
\begin{figure*}[t]
	\centering
    \begin{overpic}[width=\columnwidth]{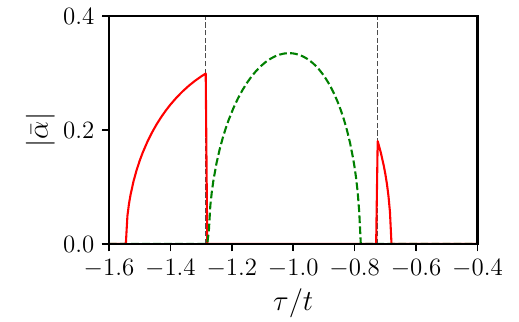}\put(0,55){a)}\end{overpic}\vspace{0em}  
    \begin{overpic}[width=\columnwidth]{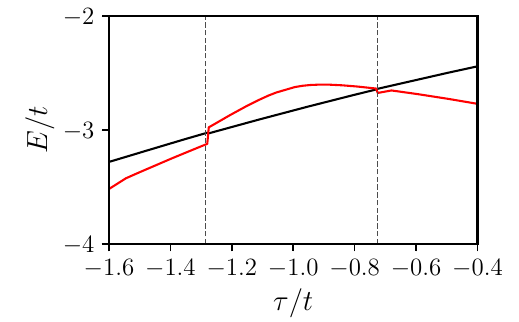}\put(0,55){b)}\end{overpic}\vspace{0em} 
    \begin{overpic}[width=\columnwidth]{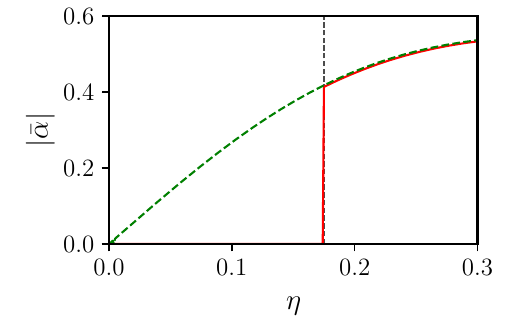}\put(0,55){c)}\end{overpic}\vspace{0em}  	
	\begin{overpic}[width=\columnwidth]{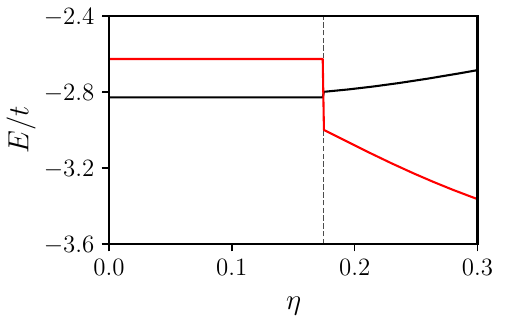}\put(0,55){d)}\end{overpic}\vspace{0em}  
	\caption{Photon condensation of square molecular plaquettes with $N_{\rm e}=2$ electrons.
    a) [c)] The photon condensate order parameter $\bar{\alpha}$ is plotted as a function of $\tau/t$ [$\eta$], by setting $\omega_{\rm c}=t$ and  $\eta=0.13$ [$\tau=-t$]. The red solid line denotes $U/t=1$, and the green dashed line refers to $U=0$.
    Furthermore, in c), the dashed gray vertical line at $\eta^*\approx0.17$ indicates the value where $\bar{\alpha}$ becomes non-zero with a jump discontinuity.
    b) [d)] Low-energy spectrum of the generic $k$-th molecule, characterized by $\hat{H}_{{\rm MF}, k}(\bar{\alpha})$, is considered with $\bar{\alpha}$ determined under the conditions $\omega_{\rm c}=t$,   $\eta=0.13$ [$\tau=-t$], and $U/t=1$. The red and black solid lines indicate the lowest eigenenergies in the subspaces ${\bf{S}}$ and ${\bf{T}}$, 
    respectively. The vertical gray dashed lines identify the critical $\tau$ values where $\bar{\alpha}$ transitions from zero. 
	}
	\label{fig:op_vs_spec_tau}
\end{figure*}
We concentrate on exploring the possible emergence of photon condensation within the context of the many-electron molecular model introduced above.
Following the approach used in Ref.~\cite{Mercurio_PRR_2024}, in the limit where $N\gg 1$, indicating the thermodynamic limit, the ground state of the complete Hamiltonian is approximated as a direct product $\ket{F} \otimes\ket{ \phi_\alpha }$. This direct product represents the quantum states of matter and light~\cite{Emary_PRL_2003, Emary_PRE_2003, chiriaco_prb_2022, passetti_prl_2023, eckhardt_commphys_2022, amelio_prb_2021}.
Specifically, $|\phi_\alpha \rangle$ denotes a coherent state that fulfills the condition $\hat a |\phi_\alpha \rangle = \alpha\sqrt{N}|\phi_\alpha \rangle$.
Photon condensation occurs when the photonic order parameter $\alpha=\bra{\phi_\alpha} \hat{a} \ket{\phi_\alpha }/\sqrt{N}$ reaches a nonzero value in the thermodynamic limit. Henceforth, we will consider $\alpha \in \mathbb{R}$ without any loss of generality.
With the thermodynamic scaling introduced above, the free electromagnetic
energy in the condensed phase is extensive,
$\omega_c\langle\hat a^\dagger\hat a\rangle\sim N$, whereas the physical
magnetic field remains intensive,
$2B_0\sqrt{N}\alpha\sim O(1)$.
Correspondingly, the Peierls phase experienced by each molecule,
$(\eta/\sqrt{N})\langle\hat a+\hat a^\dagger\rangle=2\eta\alpha$,
also remains finite in the thermodynamic limit.
When a specific $\ket{\phi_\alpha}$ is chosen, the mean-field matter Hamiltonian exclusively involves the electronic degrees of freedom:
\begin{equation}\label{mf_H}	
    \hat{\mathcal H}_{\rm{MF} }( \alpha)\equiv \langle \phi_\alpha|\hat{\mathcal H}| \phi_\alpha \rangle= \omega_{\rm c} \alpha^2 N+ \sum_{k=1}^N  \hat{H}_{{\rm MF}, k}(\alpha)~,
\end{equation}
where $\alpha$ acts on the energy of the cavity coherent field and the single Hamiltonian $\hat{H}_{{\rm MF}, k}(\alpha)=\hat{T}_{{\rm MF}, k}(\alpha)+\hat{U}_k$, where $\hat{T}_{{\rm MF}, k}(\alpha)$ is the expectation value of the kinetic term in the coherent state, and it is expressed as
\begin{equation}
    \hat{T}_{{\rm MF}, k}(\alpha)=-\sum_{j=0}^{n_{\rm s}-1}\sum_{s_z=\downarrow,\uparrow} t_{j, j+1} e^{-i 2 \eta \alpha} \hat{c}_{j  s_z k}^{\dagger} \hat{c}_{j+1 s_z  k}+\text{H.c.}~,
\end{equation}
and  $\hat{U}_k$ is the invariant repulsive Hubbard interaction.
Since the Zeeman coupling is neglected, the cavity field does not
affect the spin degrees of freedom, and
$\hat H_{{\rm MF},k}(\alpha)$ still commutes with $\hat S_z$ and
$\hat S^2$. Therefore, also for $\alpha\neq0$, the Hilbert space of
each molecule can be decomposed into subspaces with fixed total spin
$S$ and spin projection $S_z$.
Here, the mean-field molecular Hamiltonian is represented as a tight-binding model of a closed loop, which is penetrated by a transverse classical magnetic field of intensity $2B_0 \sqrt{N} \alpha $, which remains finite
in the thermodynamic limit since $B_0\propto N^{-1/2}$.

It is important to note that each molecule behaves independently of the others, also in the presence of the light field, allowing for the diagonalization of the Hamiltonian for each molecule.
Therefore, we extend the approach used in Section~\ref{sec:spectrum_nocavity}, corresponding to the case where $\alpha=0$. 
A generic many-body eigenstate of $\hat{\cal H}_{\rm MF}(\alpha)$ is denoted as 
\begin{equation}
\ket{F_\lambda(\alpha)}=\bigotimes^N_{k=1} \ket{\Psi_{\ell_k(\lambda) k}(\alpha)}~,
\end{equation}
and the associated eigenenergy is expressed as $\sum_{k=1}^N E_{\ell_k(\lambda)}(\alpha) $. The index $\ell_k(\lambda)$ specifies the eigenstate for the $k$-th molecule within the global state denoted by $\lambda$. The ground state corresponds to $\lambda = 0$ with energy $E_{0}(\alpha)$, equal for all molecules.
The order parameter $\alpha$ is obtained by evaluating the optimal value $\bar{\alpha}$ that minimizes the energy functional 
\begin{equation}\label{eq:energy-func}
{\cal E}(\alpha)=\frac{\bra{F_0(\alpha)}\hat{\cal H}_{\rm MF}(\alpha) \ket{F_0(\alpha)}}{N}=\omega_{\rm c} \alpha^2+{ E}_0(\alpha),
\end{equation}
 which leads to the following non-linear equation
\begin{equation}\label{eq:selfcons}
\partial_\alpha \mathcal{E}(\alpha)|_{\alpha=\bar{\alpha}}=
   2\omega_{\rm c} \bar\alpha + \left. \frac{\partial E_0(\alpha)}{\partial \alpha}\right|_{\alpha = \bar\alpha} = 0~.
\end{equation}
When the minimization results in $\bar\alpha \neq 0$, the system spontaneously breaks the time-reversal symmetry (TRS), leading to the initiation of a circulating current within each molecule, which produces a persistent magnetic field in the cavity.

We emphasize that the terminology photon condensation refers to the
macroscopic coherent occupation of the transverse cavity mode,
$\langle\hat a\rangle=\sqrt{N}\,\bar\alpha$, in the Coulomb gauge adopted
here. 
The ordered phase also admits a direct
characterization in terms of gauge-invariant physical observables.
Indeed, since
$\langle\hat{\bm B}\rangle=2B_0\sqrt{N}\,\bar\alpha\,\hat{\bm z}$,
a finite $\bar\alpha$ corresponds to a finite magnetic field and is
accompanied by a finite circulating orbital current, or equivalently a finite
orbital magnetization, in each plaquette. The two solutions
$\pm\bar\alpha$ therefore correspond to opposite magnetic fields and
magnetizations and are related by time reversal. In this sense, photon
condensation in the present system is equivalently a magnetostatic
instability accompanied by spontaneous time-reversal-symmetry breaking.

\subsection{Two-electron case}
\begin{figure}[t]
	\centering
	\includegraphics[width=\columnwidth]{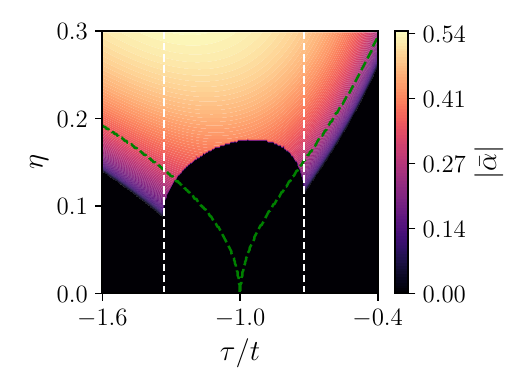}
	\caption{Photonic order parameter $|\bar\alpha|$ as a function of $\tau/t$ and the light-matter coupling strength $\eta$, with $N_{\rm e}=2$ electrons and setting $\omega_{\rm c}=t$.
    The color map represents the case with $U/t = 1$. The green dashed line indicates the critical values of the microscopic parameters $\tau$ and $\eta$, where the phase transition occurs without the influence of electron-electron interaction $U$. The vertical white dashed lines at $\tau_1\approx -1.33 t$ and $\tau_2\approx -0.72t$ denote the range $[\tau_1,\tau_2]$, within which the ground state for $\eta=0$ is a degenerate triplet.
	}	
	\label{fig:op_tau_vs_eta}  
\end{figure}
By solving Eq.~\eqref{eq:selfcons}, Fig.~\ref{fig:op_vs_spec_tau}~a) shows the order parameter $\bar{\alpha}$ as a function of $\tau/t$,  by setting $\omega_{\rm c} = t$ and $\eta=0.13$, with $U=t$ (solid red line) and $U=0$ (green dashed line). Without the repulsive electron-electron interaction, there is a finite region of $\tau$, around $\tau=-t$, where the order parameter is non-zero, and behaves as a continuous dome-shaped curve of the hopping parameter $\tau$. The condition where $|\bar{\alpha}|$ reaches its highest value arises when $\tau=-t$. For $U=0$ and $\eta=0$, this corresponds to the value of $\tau$ at which the lowest excitation energy above the ground state becomes zero, as shown in Fig.~\ref{fig:spec_U1_vs_U0}~a).
This condition is associated with a dramatically heightened paramagnetic response, known as the Van Vleck mechanism, and it explains the occurrence of a continuous photon condensation, shown in Fig.~\ref{fig:op_vs_spec_tau}~a).
Activating the Hubbard interaction $U$ results in a significant alteration of the order parameter $\alpha$. The solid red line shown in Fig.~\ref{fig:op_vs_spec_tau}~a) considers the case with $U=t$. Here, the order parameter does not follow a dome-like curve. Instead, in correspondence of two distinct values, $\tau_1^* <-t$ and $\tau_2^* < -t$ (vertical gray dashed lines), the cavity mean-field suddenly acquires a non-vanishing expectation value. At these discontinuities, the photon condensation behaves as a first-order phase transition. 
Fig.~\ref{fig:op_vs_spec_tau}~b) shows the lowest eigenenergy of the singlet subspace ${\bf{S}}$ (red solid line) and triplet subspace ${\bf{T}}$ (black solid line), as a function of $\tau/t$ and setting $\omega_{\rm c}=t$, $\eta=0.13$, and $U=t$. At the values $\tau_{{\rm c} 1}$ and $\tau_{{\rm c} 2}$, $\tau_{{\rm c} 1}<\tau_{{\rm c} 2}$, where the order parameter vanishes smoothly, 
the  second derivative of the ground state energy presents a discontinuity with respect to the hopping parameter $\tau$, 
consistently with the nature of the second-order phase transition. 
Furthermore, at values $\tau_{1}^*$ and $\tau_{2}^*$, the occurrence of a first-order phase transition is signaled by the crossing of the lowest eigenenergies within the singlet and triplet subspaces. Finally, we observe that when the system is in the photon condensate state, the matter ground state is always a singlet. 
We now apply a similar analysis by using the light-matter coupling strength $\eta$ as a tunable parameter.
Fig.~\ref{fig:op_vs_spec_tau}~c) shows the order parameter as a function of $\eta$, by setting $\omega_{\rm c}=t$ and $\tau=-t$, with $U=0$ (green dashed line) and $U=t$ (red solid line).
In the absence of repulsive interaction, photon condensation manifests as a second-order phase transition. When $\tau=-t$, the ground state of the matter system becomes gapless,
and the critical light-matter coupling is vanishing $\eta_{\rm c}=0^+$.
As discussed in Ref.~\cite{Mercurio_PRR_2024}, this behavior is attributed to a divergent paramagnetic response of the molecular system.
%
When the repulsive interaction is activated at $U=t$, photon condensation emerges as a first-order phase transition. At $\eta^* \approx 0.17$, denoted by the vertical gray dashed line in Fig.~\ref{fig:op_vs_spec_tau}~c), the order parameter $\alpha$ assumes a finite value, exhibiting a step-like behavior. In correspondence, Fig.~\ref{fig:op_vs_spec_tau}~d) shows the lowest eigenenergy of the singlet (red solid line) and triplet (black solid line) subspaces as a function of $\eta$, setting $\omega_{\rm{c}}=t$, $\tau=-t$, and $U=t$. At the value $\eta^* \approx 0.17$, associated with a first-order phase transition, there is a crossing of the lowest energies of the two allowed total spin sectors.  Here, the photon-condensation phase transition occurs, and the ground state energy lies within the singlet subspace ${\bf{S}}$. 
Moreover, for $\eta\ge\eta^*$, the energy levels become dependent on $\eta$ since $\bar{\alpha}\neq 0$.
The color map shown in Fig.~\ref{fig:op_tau_vs_eta} gives a more general overview of the impact of the repulsive electron-electron interaction on a system of molecular plaquettes with $N_{\rm e}=2$ electrons. Here, we illustrate the photonic order parameter as a function of the hopping parameter $\tau/t$ and the light-matter coupling strength $\eta$, with fixed values $\omega_{\rm c}=t$ and $U=t$. 
The green dashed line represents the critical points of the parameter set $\tau$ and $\eta$ at which the phase transition occurs without electron-electron interaction $U$. The sharp boundary between the normal phase and the photon condensate state in the central region of the figure indicates a first-order phase transition. Specifically, this occurs in the range where the hopping parameter satisfies $\tau_1<\tau<\tau_2$, with $\tau_1\approx -1.33 t$ and $\tau_2\approx -0.72t$. We recall that within this interval, in the absence of light-matter coupling, the ground state of the system with $U > 0$ is a degenerate triplet. On the other hand, outside this range of $\tau$, in conditions where the electromagnetic field of the cavity is absent, the global minimal eigenenergy is a singlet, and it is visible that photon condensation occurs as a second-order phase transition at a critical $\eta$. In this instance, the electron-electron interaction only introduces a renormalization of $\eta_{\rm c}$ with respect to the non-interacting case. 
\subsection{Three-electron case}
 \begin{figure}[t]
	\centering
	\includegraphics[width=\columnwidth]{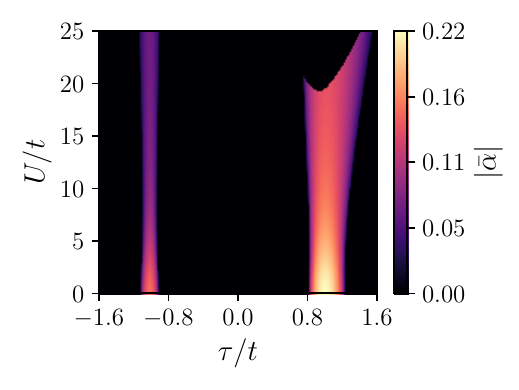}
	\caption{Photon condensation of square molecular plaquettes with $N_{\rm e}=3$ electrons.
     The photon condensate order parameter $|\bar{\alpha}|$ is plotted as a function of $\tau/t$ and $U/t$, by setting $\omega_{\rm c}=t$ and  $\eta=0.12$. The sharp boundary between the normal and photon condensate phase in the upper right corner signals a first-order phase transition.}
	\label{fig:op_vs_spec_tau_ne3}
\end{figure}
In this Section, we extend our analysis to the case where each molecule contains $N_{\rm e}=3$ electrons. We will observe that qualitatively this situation resembles the case examined previously with $N_{\rm e}=2$ electrons.
Fig.~\ref{fig:op_vs_spec_tau_ne3} shows the order parameter $|\bar{\alpha}|$, obtained by solving Eq.~\eqref{eq:selfcons}, as a function of $\tau/t$ and the electron-electron energy $U/t$, at fixed value of the light-matter coupling strength $\eta = 0.12$. We observe that in the absence of the repulsive electron-electron interaction, there are two isolated finite intervals of $\tau$, around $\tau=-t$ and $\tau=t$, where the order parameter acquires a non-zero value, and the transition behaves as a continuous phase transition. 
At $\tau = - t$ ($\tau=t$), the energy gap $\epsilon_1-\epsilon_0$ ($\epsilon_2-\epsilon_1$) between the ground state and the first excited state of the doublet subspace reduces to zero, originating a strong paramagnetic response. When $U > 0$, in the $\tau < 0$ region, the photon condensation can occur always as a second-order phase transition. In turn, for $\tau > 0$, in the presence of the $U>0$, the transition is continuous provided the electron-electron energy $U < \bar U \approx 18.6t$. When $U > \bar U$, there is a range of $\tau$ around $\tau = t$, where photon condensation occurs as a first-order phase transition. This situation strictly resembles the phenomenology which in the two-electron regime appears in the vicinity of $\tau = -t$ when the electron-electron repulsion is included, but contrary to the scenario with $N_{\rm e}=2$, in the three-electron regime the first-order phase transition is possible solely when the repulsive interaction $U$ is sufficiently large to induce level crossing between the lowest eigenenergies of the doublet and quadruplet subspaces. 
\subsection{Four-electron case}
\begin{figure}[t]
	\centering
	\includegraphics[width=\columnwidth]{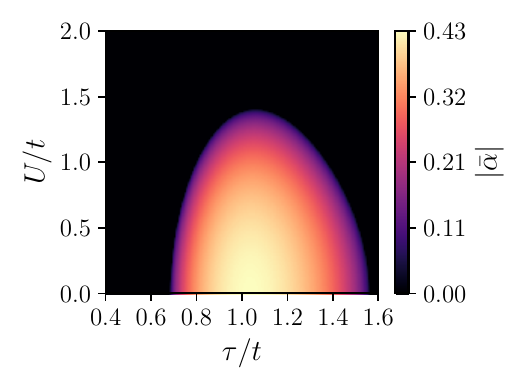}   
	\caption{Photon condensation of square molecular plaquettes with $N_{\rm e}=4$ electrons.
    The photon condensate order parameter $|\bar{\alpha}|$ is plotted as a function of $\tau/t$ and $U/t$, by setting $\omega_{\rm c}=t$ and  $\eta=0.12$. At half-filling, the transition is always of the second-order.}
	\label{fig:op_vs_spec_tau_ne4}
\end{figure}
To conclude this Section, we consider the case where each square molecular plaquette contains $N_{\rm e}=4$ electrons, representing the half-filling condition.
In contrast to previous filling conditions, here, when the cavity photon field is absent, regardless of the value of $\tau$ and for finite $U$, the ground state of the system is always linked to the singlet subspace, as visible in Fig.~\ref{fig:spec_comp}~c).
The lack of lowest energy level crossings between subspaces having different total spins clarifies why photon condensation exhibits the characteristics of a second-order phase transition, similarly to the scenarios when $N_{\rm e} = 2,3$ and the repulsive interaction $U$ is disregarded. In fact, Fig.~\ref{fig:op_vs_spec_tau_ne4} shows that photon condensation always occurs as a second-order phase transition, and the effect of the electron-electron repulsion consists only of a renormalization of the parameters at which the transition occurs. In particular, keeping $\omega_{\rm c}=t$ and $\eta=0.12$, our numerical analysis shows photon condensation is suppressed when the electron-electron repulsive interaction exceeds approximately $U_{\rm c}\approx 1.4 t$ at $\tau=t$. 
%
%
This behavior is reminiscent of the suppression of charge fluctuations in the half-filled Hubbard model in the large-$U$ regime~\cite{Arovas_AR_2022, Jeckelmann_APS_2000}, where the strong electron-electron repulsion suppresses the orbital current response and eventually inhibits photon condensation.
%
%
\section{Photon Condensation and Van Vleck Paramagnetism}
\label{sec:van_vleck}

\begin{figure*}[t]
	\centering
    \vspace{0em}
    	\begin{overpic}[width=1.9\columnwidth]{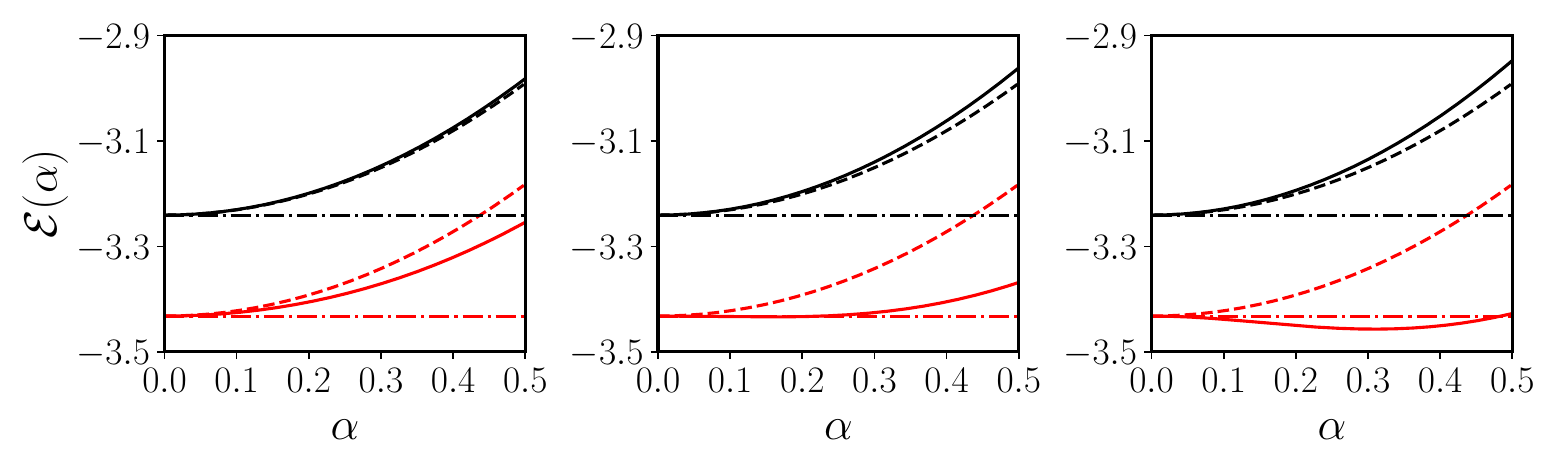}\put(11.5,25){a)}\put(43,25){b)}\put(75,25){c)}\end{overpic}
        \vspace{0em}
        \begin{overpic}[width=1.9\columnwidth]{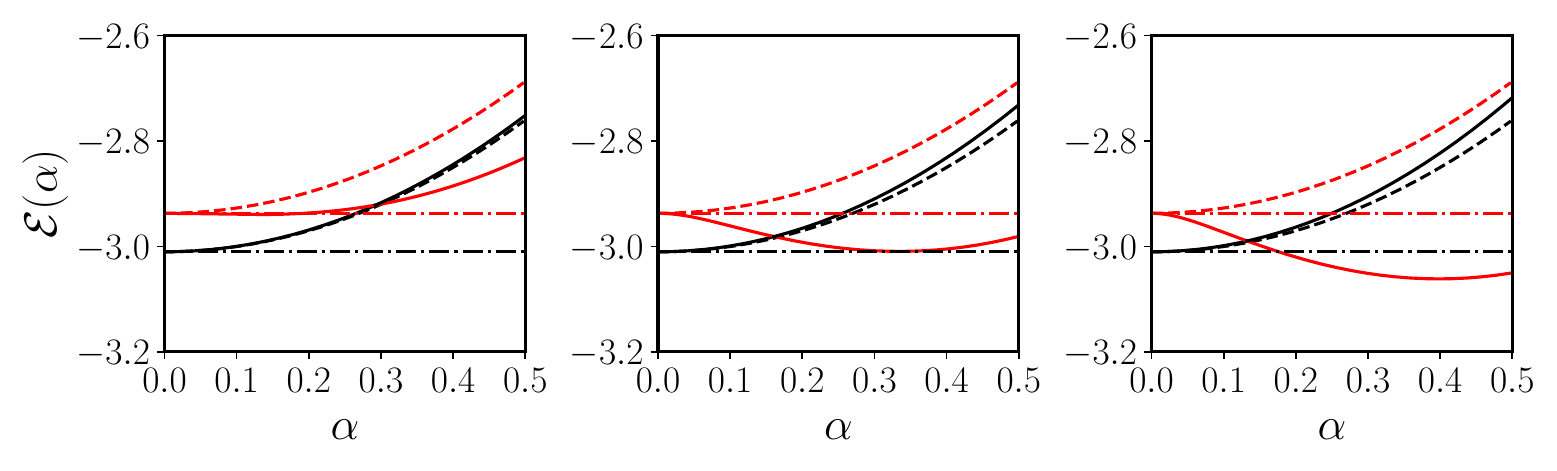}\put(11.5,25){d)}\put(43,25){e)}\put(75,25){f)}\end{overpic}
	\caption{
    The energy functionals $\mathcal{E}^{{\bf{T}}}(\alpha)$ (black solid line) and $\mathcal{E}^{{\bf{S}}}(\alpha)$ (red solid line), each plotted as a function of the photon mean field $\alpha$, with $U=t$ and $N_{\rm e}=2$.  In each panel, the dashed lines indicate $\mathcal{E}^{{\bf{T}}}(0)+\omega_{\rm c} \alpha^2=\epsilon_0+\epsilon_1+\omega_{\rm c} \alpha^2$ (black) and $\mathcal{E}^{{\bf{S}}}(0)+\omega_{\rm c} \alpha^2$ (red), and the horizontal dash-dotted lines refer to  $\mathcal{E}^{{\bf{T}}}(0) = \epsilon_0+\epsilon_1$ (black) and $\mathcal{E}^{{\bf{S}}}(0)$ (red).
    For case a), we set $\eta= 0.08$; in case b), $\eta= \eta_c \approx 0.14$; and in case c), we use $\eta=0.17$, with a value of $\tau/t = -1.55$, which falls outside the interval $[\tau_1,\tau_2]$. While, for case d), $\eta= 0.08$; for case e), $\eta= \eta^* \approx 0.14$; and for case f), $\eta=0.17$, with $\tau/t = -1.25$ lying within the interval $[\tau_1,\tau_2]$.
    }
	\label{fig:gs_energies_alpha}
\end{figure*}

Here, we discuss the underlying mechanism responsible for the dual nature of the phase transition. To this end, we focus on the two-electron regime and report the ground state energies of the light–matter system, evaluated separately in the two subspaces ${\bf{S}}$ and ${\bf{T}}$. For each subspace, the energy functional is expressed as
\begin{subequations}\label{eq:st_energy}
    \begin{align}
            \mathcal{E}^{{\bf{S}}}(\alpha) &\equiv \min_{\ket {\Psi'(\alpha)} \in {\bf{S}}} \bra{\Psi'(\alpha)}      
        \hat{H}_{{\rm{MF}},k}(\alpha) \ket{\Psi'(\alpha)}+\omega_{\rm c} \alpha^2 =\nonumber \\
        &= E_0^{\bf{S}}(\alpha) + \omega_{\rm c} \alpha^2\\
        \mathcal{E}^{{\bf{T}}}(\alpha) &\equiv \min_{\ket{\Psi(\alpha)} \in {\bf{T}}} \bra{\Psi(\alpha)}    
        \hat{H}_{{\rm{MF}},k}(\alpha) \ket{\Psi(\alpha)}+\omega_{\rm c} \alpha^2=\nonumber \\
        &=\min_{\ket {\Psi(\alpha)} \in {\bf{T}}} \bra{\Psi(\alpha)}  
        \hat{T}_{{\rm{MF}},k}(\alpha) \ket{\Psi(\alpha)}+\omega_{\rm c} \alpha^2=\nonumber\\
        &= E_0^{\bf{T}}(\alpha) + \omega_{\rm c} \alpha^2~,
    \end{align}
\end{subequations}
where the lowest eigenenergy remains identical for each square molecular plaquette, making the molecular index $k$ irrelevant.
The concavity of $E_0^{{\bf{S}}/{\bf{T}}}(\alpha)$ at the point $\alpha = 0$, for given values of $\eta$ and $\tau$ is captured by the magnetic susceptibility $\chi_{\rm M}^{{\bf{S}}/{\bf{T}}}= -\frac{1}{2}\partial^2_\alpha [E_0^{{\bf{S}}/{\bf{T}}}(\alpha)]|_{\alpha = 0}$, whose sign allows us to distinguish the magnetic characteristics of the ground state within each subspace as either paramagnetic or diamagnetic. In particular, the nature is identified as paramagnetic (diamagnetic) when the concavity is negative (positive) due to the energetic favorability of penetration by a magnetic field. By performing a series expansion of Eq.~\eqref{eq:st_energy} around $\alpha = 0$ up to second order in $\alpha$, we obtain
\begin{equation}
    \mathcal{E}^{{\bf{S}}/{\bf{T}}}(\alpha) = E_0^{{\bf{S}}/{\bf{T}}}(0) + \Big(\omega_{\rm c}-\chi_{{\rm M}}^{{\bf{S}}/{\bf{T}}}\Big) \alpha^2~.
\end{equation}
When $\chi_{{\rm M}}^{{\bf{S}}/{\bf{T}}} > \omega_{\rm{c}}$, the energy decreases in the presence of a finite cavity field. Fulfilling this condition demands a positive magnetic susceptibility with a sufficiently large magnitude.
Fig.~\ref{fig:gs_energies_alpha} shows the representative cases that clarify the characteristics of the phase transition, in the presence of the repulsive electron-electron interaction.
In the upper panels of Fig.~\ref{fig:gs_energies_alpha}, the hopping parameter is set to $\tau/t = -1.55$ with $U=t$, explored in three values of $\eta$. 
In a), we set $\eta = 0.08<\eta_{\rm c}$, in b), $\eta = \eta_c$, and in c), $\eta = 0.17>\eta_{\rm c}$.
This particular value of $\tau$ lies outside the interval $[\tau_1,\tau_2]$. In fact, when the cavity field $\alpha$ is absent, the ground state belongs to the singlet subspace,  $\mathcal{E}^{{\bf{S}}}(0) < \mathcal{E}^{{\bf{T}}}(0)$.
Here, the fundamental energies in the absence of the light field are marked by the horizontal dot-dashed lines.
 Considering $\eta$ as a parameter, the photon condensation manifests itself as a second-order phase transition, taking place at a critical value $\eta_c \approx 0.14$.
The matter sector displays a paramagnetic nature, resulting in a decrease in the concavity at $\alpha=0$ of $\mathcal{E}(\alpha)$ as $\eta$ increases, ultimately leading to a sign change at the critical point $\eta_{\rm c}$, as illustrated in Fig.~\ref{fig:gs_energies_alpha}~b). Consequently, when $\eta$ exceeds $\eta_{\rm c}$, the global minimum of $\mathcal{E}(\alpha)$ emerges at a finite $\alpha$, as shown in Fig.~\ref{fig:gs_energies_alpha}~c).
For these three specific values of $\eta$, the global minimum energy always pertains to the singlet subspace ${\bf{S}}$.
In the bottom panels of Fig.~\ref{fig:gs_energies_alpha}, the hopping parameter is set at $\tau/t = -1.25$ with $U=t$, and we consider three distinct values of $\eta$. 
In e) we choose $\eta = 0.08<\eta^*$, in d), $\eta = \eta^*$, and in f), $\eta = 0.17>\eta^*$.
Here, the hopping parameter $\tau$ lies within the interval $[\tau_1,\tau_2]$. 
In this case, when the cavity field $\alpha$ is switched off, the ground state lies in the triplet subspace,  $\mathcal{E}^{{\bf{T}}}(0) < \mathcal{E}^{{\bf{S}}}(0)$.
%
%
Comparing $\mathcal{E}^{{\bf{S}}}(\alpha)$ (red solid line) and $\mathcal{E}^{{\bf{S}}}(0)+\omega_{\rm c}\alpha^2$, (red dashed line), we notice a decrease in concavity at $\alpha=0$ as $\eta$ increases, which indicates a paramagnetic character.
Employing a similar analysis for the  triplet subspace, denoted by black lines, it is observed that the concavity of $\mathcal{E}^{{\bf{T}}}(\alpha)$ increases with an increase in $\eta$, indicating negative magnetic susceptibility and diamagnetic character of the triplet sector. 
%
This contrasting magnetic response explains the change in the nature of the phase transition within the region $[\tau_1, \tau_2]$. In fact, when $\eta < \eta^*$ ($\eta^*\approx 0.14$ when $\tau = -1.25t)$, the global energy minimum corresponds to $\alpha = 0$ and is linked to the triplet subspace. In turn, for $\eta = \eta^* + 0^+$, the energy minimum at $\alpha = 0$ suddenly becomes local, the global eigenenergy of the full light-matter system being associated with a finite photonic displacement in the cavity, when $\eta > \eta^*$, the photon condensation appears as a first-order phase transition. Once the transition takes place, the ground state of the matter system is linked to the singlet subspace. 
Here, when photon condensation occurs, the matter ground state switches from the triplet to the singlet sector, in which the Hubbard interaction contributes to the energy.
In summary, this analysis shows that when the electron occupation number is $N_{\rm e}=2$, the repulsive interaction between electrons can significantly alter the character of photon condensation, provided the microscopic parameters lie within a certain range. 
This discussion can also be extended to the situation with $N_{\rm e}=3$ electrons per molecule, in which photon condensation can emerge either as a first-order or as a second-order phase transition.

We finally comment on some aspects related to the cavity description and experimental realization. For second-order photon-condensation transitions, multimode effects can be addressed by analyzing the stability of the normal phase with respect to the different electromagnetic modes~\cite{Andolina_APS_2020}. Once condensation occurs, however, the electronic ground state is modified, and the role of the remaining modes should in general be reconsidered. Moreover, such a normal-phase stability analysis is not sufficient for the first-order transitions found here. Additional cavity modes may therefore modify the phase boundaries, and a complete multimode treatment would require specifying a particular cavity realization and its full mode structure.

Effective Hubbard-type descriptions of conjugated molecules typically involve electronic energy scales in the eV range, with $U/t$ of order unity or larger. Although we consider $\omega_{\rm c}\sim t$ in our calculations, motivated by the optical-cavity regime, this is not required for the instability, which does not rely on resonance between the cavity mode and an electronic excitation. In fact, for a given magnetic response, a lower cavity energy $\omega_{\rm c}$ reduces the electromagnetic energy cost and
therefore favors the instability.
From an experimental perspective, the magnetic coupling $\eta$ is controlled by the vacuum magnetic flux threading the molecular plaquette. The values considered here, typically $\eta\sim0.1$, should be regarded as representative
theoretical values. The critical coupling can, however, be reduced near magnetically active quasi-degeneracies, where the Van Vleck orbital susceptibility is enhanced. Superconducting nanoconstrictions and lumped-element resonators have recently demonstrated strong confinement of vacuum magnetic fields in molecular spin systems ~\cite{gimeno_acsnano_2020,rubin_arxiv_2025}. Although these experiments involve Zeeman rather than orbital coupling, they demonstrate the possibility of strongly concentrating the vacuum magnetic field near molecular systems.
For the orbital coupling examined here, the limited size of conventional molecules, and thus the correspondingly small vacuum flux $B_0 A_p$, constitutes an experimental limitation.

\section{Polaritons}
\label{sec:polaritonic_spectrum}

Polaritons are novel hybrid light-matter states, arising as collective excitations above the mean-field ground state. These excitations result from the interaction between photonic modes and collective matter excitations.
%
To find polaritons in our system, we need to study Gaussian fluctuations around the mean-field state.
%
%
Starting from the full Hamiltonian expressed in Eq.~\eqref{eq:Hfull},
 to derive the polaritonic spectrum we replace $\hat{a} \rightarrow \bar{\alpha} \sqrt N+\delta \hat a$  and  $\hat{a}^\dagger \rightarrow \bar{\alpha} \sqrt N+\delta \hat a^\dagger$, where $\delta \hat{a}$ and $\delta \hat{a}^\dagger$ are zero-average fluctuation operators on top of the mean-field solution $\bar\alpha\sqrt{N}$. We introduce the collective bright mode creation operator $\hat{b}_\ell^\dagger$, defined as
\begin{equation}
	\hat{b}_\ell^\dagger \equiv \frac{1}{\sqrt{N}} \sum_{k=1}^N |\Psi_{\ell k}(  \bar\alpha)\rangle\langle \Psi_{0 k}(\bar\alpha)|~,
\end{equation}
with $\ell > 0$, where the action of the bright mode operator on the many-body Hilbert space is to promote the electronic state of the $k$-th molecule in the ground state $\ket{\Psi_{0k}(\bar\alpha)}$ to the $\ell$-th excited state $\ket{\Psi_{\ell k}(\bar\alpha)}$, the associated transition energy being $E_\ell(\bar\alpha) - E_0(\bar\alpha)$. 
In the thermodynamic limit $N \to \infty$, the bright mode creation operators effectively satisfy bosonic commutation relations, namely $[\hat{b}_\ell, \hat b_m^\dagger] \approx \delta_{\ell m}$, details are provided in Appendix~\ref{app:bosonization}.
Here, we focus on small fluctuations over the mean-field solution. 
For this aim, the full Hamiltonian is expanded up to the second-order with respect to the photonic fluctuations $\delta \hat{a}$ and  $\delta \hat{a}^\dagger$ and the bright mode operators $\hat{b}_\ell$ and  $\hat{b}^\dagger_\ell$. This yields the polaritonic Hamiltonian, which characterizes the lowest excited states
\begin{equation}
\begin{aligned}
	\hat{\mathcal{H}}_{\rm{pol}} & =\omega_{\rm c} \delta \hat a^\dagger \delta \hat{a}+\sum_{\ell}[E_\ell(\bar \alpha)-E_0(\bar \alpha)] \hat b_\ell^{\dagger} \hat{b}_\ell \\
	& +\frac{1}{2}(\delta \hat{a}+\delta \hat a^\dagger) \sum_{\ell}[\mathcal{M}_{\rm{p}}^{\ell, 0}(\bar \alpha) \hat{b}_\ell^{\dagger}+\mathcal{M}_{\rm{p}}^{0, \ell}(\bar\alpha) \hat{b}_\ell] \\
	& +\frac{1}{8} (\delta\hat{a}+\delta \hat a^\dagger)^2 \mathcal{M}_{\rm d}^{0, 0}(\bar \alpha)
\end{aligned}
\end{equation}
where the optimal value $\bar\alpha$ is determined by the variational approach shown in Sec.~\ref{sec:mf_theory}, and $\mathcal{M}_{{\rm p}/{\rm d}}^{\ell, m}(\bar{\alpha})=\sum_{k=1}^N \langle\Psi_{\ell k}(\bar{\alpha})| \hat{\mathcal{M}}_{{\rm p}/{\rm d}}(\bar{\alpha})|\Psi_{m k}(\bar{\alpha})\rangle$, being
\begin{equation}\label{eq:par_magnetization}
	\hat{\mathcal M}_{\rm p}(\alpha)=\frac{1}{N}\sum_{k=1}^N \frac{ \partial\hat H_{{\rm{MF}},k}(\alpha)}{\partial\alpha} =\frac{1}{N}\sum_{k=1}^N \frac{ \partial\hat T_{{\rm{MF}},k}(\alpha)}{\partial\alpha}~,
\end{equation}
\begin{equation}
	\hat{\mathcal M}_{\rm d}(\alpha)= \frac{\partial\hat{\cal M}_{\rm p}(\alpha)}{\partial\alpha}~,
\end{equation}
the paramagnetic and diamagnetic contributions. Notice the paramagnetic component is proportional to the magnetization operator
\begin{equation}
\begin{aligned}
    \hat M_z(\alpha) &= -\sqrt{N} \hat{\mathcal M}_{\rm p}(\alpha) /2B_0=-\frac{\eta}{B_0 \sqrt{N}} \times\\ 
    &\times \sum_{k, j} \sum_{s_z = \downarrow, \uparrow} (i t_{j,j+1} e^{-i 2 \eta\alpha} \hat{c}_{j s_z k}^{\dagger} \hat{c}_{j+1 s_z k}+ \text{H.c.})\\
\end{aligned}
\end{equation}

The fluctuations $\delta\hat a$ and $\hat b_\ell$ entering $\hat H_{\rm pol}$ provide subextensive corrections around the mean-field solution. In particular, since $\hat a/\sqrt{N}=\bar\alpha+\delta\hat a/\sqrt{N}$, fluctuations of the intensive cavity-field amplitude are suppressed by $N^{-1/2}$. 
Accordingly, the Gaussian corrections to the ground-state energy vanish at the level of the energy density in the thermodynamic
limit~\cite{Wang-Hioe_PRA_1973,hepp_pra_1973,pellegrino_prb_2014}.

The polaritonic Hamiltonian is quadratic in the bosonic fields, by diagonalizing the corresponding Hopfield matrix, one directly obtains the associated spectrum~\cite{Hopfield_pr_1958}, all details are provided in Appendix~\ref{app:bosonization}.
%

\begin{figure}[t]
	\centering
	\vspace{0em}  
	\begin{overpic}[width=\columnwidth]{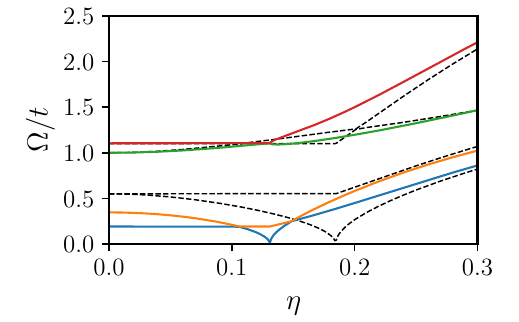}\put(0,55){a)}\end{overpic}\vspace{0em} 
    \begin{overpic}[width=\columnwidth]{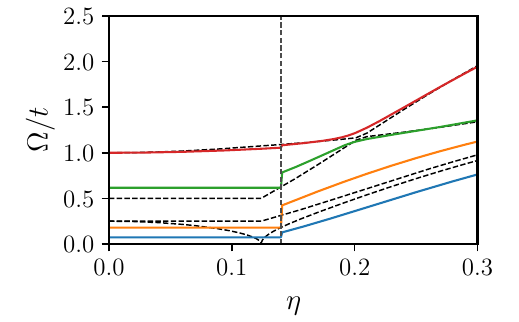}\put(0,55){b)}\end{overpic}
	\caption{Four lowest polaritonic energies as a function of $\eta$, for $U = 0$ (black dashed lines) and $U = t$. In a), the hopping parameter is $\tau/t = -1.55$, and there is a second-order phase transition for both $U=0$ and $U=t$. In b), $\tau/t = -1.25$. In the non-interacting case, the phase transition is of the second-order. While for $U = t$, the vertical dashed line marks the discontinuity in the polaritonic energies at $\eta = \eta^*$, signaling a first-order phase transition. 
	}\label{fig:polaritons}
\end{figure}

Here, as a representative example, we consider the case with $N_{\rm e} = 2$ electrons per molecule.
Fig.~\ref{fig:polaritons} shows the polaritonic spectrum, as a function of $\eta$ at different inequivalent hoppings $\tau$. 
Specifically, Fig.~\ref{fig:polaritons}~a) displays the four lowest polaritonic energies as a function of the light-matter coupling strength $\eta$, in the absence of electron-electron repulsion (black dashed lines), and with $U/t = 1$ (colored solid lines), setting $\tau = -1.55t$. This case corresponds to the top panels shown in Fig.~\ref{fig:gs_energies_alpha}. Here, regardless of the value of $\eta$, the mean-field solution indicates that the matter sector of the ground states resides within the singlet subspace ${\bf{S}}$, and the photon condensation shows characteristics of a second-order phase transition. 
In Fig.~\ref{fig:polaritons}~a), when $\eta$ reaches the critical threshold of $\eta_{\rm c}\approx0.14$, the softening of the lowest polariton mode is observed,  signaling that the photon condensation is a second-order phase transition. This phenomenon is observed in both the interacting and non-interacting cases. In particular, at the hopping parameter $\tau=-1.55 t$, the presence of the repulsive interaction leads to a renormalization of the critical value $\eta_{\rm c}$.
Fig.~\ref{fig:polaritons}~b) shows the four lowest polaritonic energy levels as a function of the light-matter interaction strength $\eta$, with $\tau$ fixed at $-1.25t$. 
The black dashed lines represent the case without electron-electron interaction, while the colored solid lines correspond to a case where $U/t = 1$.
For $U=0$, the polaritonic softening observed at $\eta_{\rm c} \approx 0.17$ indicates a second-order phase transition. In contrast, when $U = t$, the parameter $\tau = -1.25t$ falls within the interval $[\tau_1, \tau_2]$. At $\eta = 0$, the ground state of the matter system is a triplet degenerate state, with its eigen-energy unaffected by the value of $U$.
From Fig.~\ref{fig:op_tau_vs_eta}, we note that, in this configuration, the transition is characterized as first-order. This behavior can be observed in the low-energy polaritonic spectrum shown in Fig.~\ref{fig:polaritons} b), where all excitation energies display a jump associated with the first-order phase transition, marked by the vertical dashed line at $\eta = \eta^* \approx 0.14$.

In a more realistic description, cavity losses, as well as relaxation and dephasing of matter excitations, result in finite linewidths of the polaritonic modes. For a second-order transition, the softening of the lowest polariton remains spectroscopically distinguishable as long as its linewidth $\Gamma_-$  is much smaller compared to its bare frequency $\Omega_-$, i.e., $\Gamma_- \ll \Omega_-$. However, sufficiently close to the critical point, where $\Omega_- \to 0$, this condition will eventually cease to be satisfied. For a first-order transition, the abrupt change in polaritonic energies remains observable as long as the associated linewidths are smaller than the characteristic energy jump.

\section{Conclusions}
\label{sec:conclusions}


In this work, we investigated how electron–electron repulsion affects both the onset and the nature of the photon condensation phase transition in a system composed of many coplanar molecules coupled to a single quantized magnetic mode.
Each molecule is modeled as a plaquette with $n_{\rm s}=4$ sites, where each hosts $N_{\rm e}$ spinful electrons, with $1 < N_{\rm e} \leq n_{\rm s}$.
For multiple electrons per molecule,  our results demonstrate that photon condensation can occur even in the presence of electron–electron repulsion, which can substantially reshape the phase diagram compared to the non-interacting case. In particular, depending on the number of itinerant electrons and the microscopic parameters that enter the molecular Hamiltonian, photon condensation can manifest as either a first or a second-order phase transition.
Furthermore, we have demonstrated that the polaritonic spectrum can be used to probe the character of the phase transition. In particular, we found that polaritonic modes display either a softening or a step-like change, depending on whether the transition is continuous or discontinuous.
Possible future developments may include the investigation of reflectivity and transmission spectra in the vicinity of the phase transition~\cite{lamberto_quantum_2006}, potentially providing additional spectroscopic signatures of photon condensation, as well as the study of finite-temperature effects and Zeeman coupling on the phase diagram and polaritonic excitations.

\begin{acknowledgments}
	The authors thank G.G.N. Angilella, F. Bonasera, and V. Varrica, for their insightful comments and constructive feedback throughout various stages of this work.
    M. Parisi and E.P. thank the PNRR MUR project PE0000023-NQSTI.
    E.P. acknowledges support from COST Action CA21144 superqumap. 
    F.M.D.P. acknowledges support from the project PRIN 2022 - 2022XK5CPX (PE3) SoS-QuBa - ``Solid State Quantum Batteries: Characterization and Optimization". 
    G.F. thanks for the support ICSC - Centro Nazionale di Ricerca in High-Performance Computing, Big Data and Quantum Computing under project E63C22001000006, and Universit\`a degli Studi di Catania, project TCMQI PIACERI 2024/2026. G.M.A acknowledges funding from the European Union’s Horizon 2020 research and innovation programme under the Marie Sklodowska-Curie (Grant Agreement No. 101146870–COMPASS).
\end{acknowledgments}

\appendix
\begin{widetext}
\section{Factorization of the many-body Hilbert space}\label{app:Hilbert_space_structure}
In this Appendix, we show how the SU$(2)$ symmetry of the Hubbard model enables a decomposition of the full Hilbert space of the single molecule in invariant spin sectors. We consider the single-molecule Hamiltonian 
\begin{equation*}
    \hat H = \hat{T} + \hat U~,
\end{equation*}
being 
\begin{align*}
    \hat{T}&=-\sum_{j=0}^{n_{\rm s}-1} \sum_{s_z=\downarrow, \uparrow}\left(t_{j, j+1} \hat{c}_{j s_z}^{\dagger} \hat{c}_{j+1 s_z}+\text { H.c. }\right)~,\\
    \hat U &= U\sum_{j=0}^{n_{\rm{s}}-1} \hat{c}_{j \uparrow}^{\dagger} \hat{c}_{j \uparrow} \hat{c}_{j \downarrow}^{\dagger} \hat{c}_{j \downarrow}~.
\end{align*}
where we omitted the molecular index $k$ for the sake of notational simplicity. The Hamiltonian $ \hat H $ includes only processes that conserve both the total spin projection along the $z$ axis and the total spin magnitude. This implies that the operators $\hat S_z$ and $\hat S^2$ commute with the Hamiltonian, i.e.: $[\hat H, \hat{S}_z] = [\hat H, \hat{S}^2] = 0$. As a consequence, the full Hilbert space of the system $\bf H$ can be decomposed as a direct sum of subspaces characterized by the quantum numbers $S_z$ and $S$. This enables us to construct a basis of simultaneous eigenstates of operators $\hat H, \hat S_z$, and $\hat S^2$, such that any eigenstate of the Hamiltonian can be characterized by its energy $E$, spin projection along $z$  $S_z$, and total spin modulus $S$. In order to determine the common eigenbasis, we first introduce the spin operators relevant to our discussion. The spin operator describing the spin projection along the $\mu = x,y,z$ axis in the lattice basis can collectively be written as
\begin{equation}
	\hat{S}_\mu= \frac{1}{2}\sum_{j, \alpha\beta} \hat c^\dagger_{j\alpha} \sigma_\mu^{\alpha\beta}\hat c_{j\beta}~,
\end{equation}
where 
\begin{equation}
	\bm{\sigma} = \left[\left(\begin{array}{ll}
		0 &1 \\
		1 & 0
	\end{array} \right),\left(\begin{array}{cc}
		0 & -i \\
		i &0
	\end{array} \right),\left(\begin{array}{cc}
		1 & 0 \\
		0 &-1
	\end{array}\right)\right]^{\rm T}~.
\end{equation}
Explicitly, the spin components read:
\begin{subequations}
	\begin{align}
		\hat S_x & =\frac{1}{2} \sum_j (\hat c^\dagger_{j\uparrow} \hat c_{j\downarrow} +\hat c^\dagger_{j\downarrow} \hat c_{j\uparrow})~,\\
		\hat S_y & = \frac{1}{2i} \sum_j (\hat c^\dagger_{j\uparrow} \hat c_{j\downarrow} -\hat c^\dagger_{j\downarrow} \hat c_{j\uparrow})~,\\
		\hat S_z &=  \frac{1}{2}\sum_j(\hat c^\dagger_{j\uparrow} \hat c_{j\uparrow} -\hat c^\dagger_{j\downarrow} \hat c_{j\downarrow})~.
	\end{align}
\end{subequations}
It is also convenient to define the spin raising and lowering operators: 
\begin{subequations}
	\begin{align}
		\hat S_{+} &= \hat{S}_x + i\hat{S}_y =\sum_j \hat c_{j \uparrow}^{\dagger} \hat c_{j \downarrow}~,\\
		\hat S_{-} &= \hat{S}_x - i\hat{S}_y =\sum_j \hat c_{j \downarrow}^{\dagger} \hat c_{j \uparrow}~,
	\end{align}
\end{subequations}
both $\hat S_{+}$ and  $\hat S_{-}$ can be expressed as linear combinations of the $\hat S_{\mu}$ operators and commute with the Hamiltonian. To compute the total spin squared operator $\hat S^2 = \hat S_x^2 + \hat S_y^2+ \hat S_z^2$, we start from the contribution from the $z$ component, which results in
\begin{equation}
	\begin{aligned}
		\hat S_z^2 &= \frac{1}{4} \sum_{i,j}  \hat S_z^i \hat S_z^j =\\
		&= \frac{1}{4}  \sum_{i,j} (\hat n_{i\uparrow} - \hat n_{i\downarrow}) (\hat n_{j\uparrow} - \hat n_{j\downarrow})=\\
		&= \frac{1}{4} \sum_{i}(\hat n_{i\uparrow} - \hat n_{i\downarrow})^2 + \frac{1}{4} \sum_{i\neq j} \hat S_z^i \hat S_z^j~,
	\end{aligned}
\end{equation}
Proceeding similarly, for the remaining components we find
\begin{equation}
	\begin{aligned}
		\hat S_x^2 &=\frac{1}{4} \sum_{i,j} (\hat c^\dagger_{i\uparrow} \hat c_{i\downarrow} +\hat c^\dagger_{i\downarrow} \hat c_{i\uparrow}) (\hat c^\dagger_{j\uparrow} \hat c_{j\downarrow} +\hat c^\dagger_{j\downarrow} \hat c_{j\uparrow}) = \\
		&= \frac{1}{4} \sum_{i}(\hat n_{i\uparrow} - \hat n_{i\downarrow})^2 + \frac{1}{4} \sum_{i\neq j} \hat S_x^i \hat S_x^j~,
	\end{aligned}
\end{equation}
and 
\begin{equation}
	\begin{aligned}
		\hat S_y^2 &=-\frac{1}{4} \sum_{i,j} (\hat c^\dagger_{i\uparrow} \hat c_{i\downarrow} -\hat c^\dagger_{i\downarrow} \hat c_{i\uparrow}) (\hat c^\dagger_{j\uparrow} \hat c_{j\downarrow} -\hat c^\dagger_{j\downarrow} \hat c_{j\uparrow}) = \\
		&= \frac{1}{4} \sum_{i}(\hat n_{i\uparrow} - \hat n_{i\downarrow})^2 - \frac{1}{4} \sum_{i\neq j} \hat S_y^i \hat S_y^j~.
	\end{aligned}
\end{equation}
Collecting all the results above, we obtain
\begin{equation}
	\hat S^2 = \frac{3}{4} \sum_{i}(\hat n_{i\uparrow} - \hat n_{i\downarrow})^2 +\frac{1}{4}\sum_{i,j} 	S_x^i \hat S_x^j +- S_y^i \hat S_y^j + 	S_z^i \hat S_z^j~.
\end{equation}
To simplify this expression, we first observe that
\begin{equation}
	S_x^i \hat S_x^j-S_y^i \hat S_y^j = 2(\hat c^\dagger_{i\uparrow} \hat c_{i\downarrow}\hat c^\dagger_{j\downarrow} \hat c_{j\uparrow} + \hat c^\dagger_{i\downarrow} \hat c_{i\uparrow}\hat c^\dagger_{j\uparrow} \hat c_{j\downarrow}) = 2(\hat S_i^+ \hat S_j^- + \hat S_i^- \hat S_j^+)~,
\end{equation}
and, finally, we write
\begin{equation}
	\hat S^2 = \frac{3}{4} \sum_{i}(\hat n_{i\uparrow} - \hat n_{i\downarrow})^2+ \sum_{i\neq j} \left[ \frac{1}{2}(\hat S_i^+ \hat S_j^- + \hat S_i^- \hat S_j^+) + \frac{1}{4} (\hat n_{i\uparrow} - \hat n_{i\downarrow})(\hat n_{j\uparrow} - \hat n_{j\downarrow})\right]~.
\end{equation}
This final expression provides the second-quantized form of the operator $\hat{S}^2$, written in terms of fermionic creation and annihilation operators in the lattice basis. We observe that the same operators can be straightforwardly written in the eigenstate basis of the kinetic energy $\hat{T}$ by means of the transformation
\begin{subequations}
\begin{align}
    \hat{c}_{j \alpha}&=\sum_\ell \varphi_{j}^\ell \hat{\varphi}_{\ell \alpha}~,\\
    \hat{c}^\dagger_{j \alpha}&=\sum_\ell \varphi_{j}^\ell \hat{\varphi}^\dagger_{\ell \alpha}~,
\end{align}  
\end{subequations}
where $\alpha = \uparrow,\downarrow$, and used that in the absence of light-matter coupling, the eigenvectors are real. Here, the operator $\hat{\varphi}^\dagger_{\ell s_z}$ creates an electron in the $\ell$-th eigenstates, $\ell = 0, \ldots, n_{\rm s} -1$, with energy $\epsilon_\ell$ and the spin projection along $z$ is $\alpha$. 
Accordingly, one can rewrite the spin operators as
\begin{equation}
\begin{aligned}
    \hat{S}_\mu &= \frac{1}{2}\sum_{\ell,m, \alpha\beta} \hat\varphi^\dagger_{\ell\alpha} \sigma_\mu^{\alpha\beta}\hat\varphi_{m\beta}\left(\sum_j \varphi_j^\ell \varphi_j^m\right)= \frac{1}{2}\sum_{\ell, \alpha, \beta} \hat\varphi^\dagger_{\ell\alpha} \sigma_\mu^{\alpha\beta}\hat\varphi_{\ell\beta}~.
\end{aligned}
\end{equation}
%

We begin by considering the two-electron sector, where the total spin of the system can be either $S=0$ (singlet) or $S=1$ (triplet). Consequently, the full Hilbert space is decomposed as $\mathbf{H}=\mathbf{T} \oplus \mathbf{S}$, where $\mathbf{T}$ denotes the triplet subspace and $\mathbf{S}$ the singlet one.
Within the triplet sector, the allowed spin projections along the $z$-axis are $S_z=-1, 0, 1$, which allows us to further decompose $\bf T$ as 
\begin{equation}
    {\bf T} =\bigoplus_{i=-{1}}^{1}{ \bf T}_i~,
\end{equation}
where ${\bf{T}}_i$ collects the triplet states with $i$ spin projection along $z$. The subspace ${\bf{T}}_1$ has dimension 6, and its states are expressed in the common eigenbasis of $\hat{H}, \hat{S}_z$ and $\hat S^2$ as
\begin{equation}
    \ket{T^{1}_{(\ell, m)}} = \hat{\varphi}^\dagger_{\ell \uparrow} \hat{\varphi}^\dagger_{m \uparrow}\ket{\emptyset}~,
\end{equation}
with $\ell \neq m$. It is easy to verify that such states are simultaneous eigenstates of $\hat H, \hat S_z, \hat S^2$, and it results
\begin{equation}
    \begin{aligned}
        \hat H \ket{T^{1}_{(\ell, m)}}  &= (\epsilon_{\ell} + \epsilon_m)\ket{T^{1}_{(\ell, m)}}~, \\
        \hat S_z \ket{T^{1}_{(\ell, m)}}  &= \ket{T^{1}_{(\ell, m)}} ~,\\
        \hat S^2 \ket{T^{1}_{(\ell, m)}}  &= 2\ket{T^{1}_{(\ell, m)}} ~.
    \end{aligned}
\end{equation}
The state $\ket{T^{1}_{(\ell, m)}}$ is constructed as a two-particle eigenstate of the kinetic energy operator $\hat{T}$, but it is also an eigenstate of the Hamiltonian $\hat{H}$, which includes the on-site electron-electron repulsion.
The state $\ket{T^{1}_{(\ell, m)}}$ denotes a two-electron configuration in which both electrons possess spin $S_z = 1/2$. Because of the Pauli exclusion principle, two electrons with the same spin $z$-component cannot simultaneously occupy the same site. Consequently, the local Hubbard interaction does not act on $\ket{T^{1}_{(\ell, m)}}$.
The triplet states with lower spin projection are obtained by acting on the triplet states in the $S_z = 1$ sector with the lowering operator $\hat S_-$:
\begin{equation}
    \hat{S}_-\ket{T^{1}_{(\ell, m)}} = (\hat{\varphi}^\dagger_{\ell \uparrow} \hat{\varphi}^\dagger_{m \downarrow} + \hat{\varphi}^\dagger_{\ell \downarrow} \hat{\varphi}^\dagger_{m \uparrow})\ket{\emptyset}~,
\end{equation}
Hence, the six normalized $S_z = 0$ states of ${\bf {T}}_0$ read
\begin{equation}
    \ket{T^{0}_{(\ell, m)}} = \frac{1}{\sqrt{2}}(\hat{\varphi}^\dagger_{\ell \uparrow} \hat{\varphi}^\dagger_{m \downarrow} + \hat{\varphi}^\dagger_{\ell \downarrow} \hat{\varphi}^\dagger_{m \uparrow})\ket{\emptyset}~.
\end{equation}
Since $[\hat H, S^{-}] = 0$,  one has $\hat H\Big(\hat S_- \ket{T^1_{(\ell, m)}}\Big) = \hat S_- \Big(\hat H\ket{T^1_{(\ell,m)}}\Big)$. It follows that $\hat H \ket{T^0_{(\ell, m)}} = (\epsilon_\ell + \epsilon_m) \ket{T^0_{(\ell, m)}}$, and we can conclude $\ket{T^0_{(\ell,m)}}$ is eigenstate of $\hat H$ with energy $\epsilon_\ell + \epsilon_m$, independent of $U$.
By applying again the lowering operator, we can also obtain the six states belonging to ${\bf {T}}_{-1}$, which can be written as 
\begin{equation}
    \hat{S}_-\ket{T^{0}_{(\ell, m)}} \equiv \ket{T^{-1}_{(\ell, m)}} = \hat{\varphi}^\dagger_{\ell \downarrow} \hat{\varphi}^\dagger_{m \downarrow}\ket{\emptyset}~,
\end{equation}
each characterized by the quantum numbers $E=\epsilon_\ell + \epsilon_m$, $S_z = -1, S = 1$. 
In conclusion, the $\bf T$ subspace of the Hilbert space is spanned by the degenerate triplets
\begin{equation}
    \Big\{\ket{T^{1}_{(\ell, m)}}, \ket{T^{0}_{(\ell, m)}}, \ket{T^{-1}_{(\ell, m)}}\Big\}~,
\end{equation}
each having an energy $\epsilon_\ell + \epsilon_m$ which is not influenced by the electron-electron interaction. 

We now focus our attention on the three-electron case. Proceeding as in the two electron regime, we exploit the SU$(2)$ symmetry of the Hubbard Hamiltonian to express the Hilbert space as a direct sum of spin sectors. In particular, the allowed total spins for a three fermions system are $S = 1/2, 3/2$. 
The subspace with total spin $S = 1/2$ ($S = 3/2$) can be further decomposed according to the spin projection along the $z$ axis, $S_z = \pm 1/2$ ($S_z = \pm 3/2, \pm 1/2$). Consequently, the Hilbert space can be written as
\begin{equation}
{\bf H} = \bigoplus_{i=-3/2}^{3/2}{\bf Q}_i \bigoplus_{i=-1/2}^{1/2}{\bf D}_i~.
\end{equation}
Clearly, the fully polarized quadruplet states with spin projection $S_z = \pm 3/2$ are unaffected by $U$, since there are no electronic configurations in which any atomic site becomes doubly occupied.
As a consequence, their eigenenergies are expressed as $\epsilon_\ell + \epsilon_m + \epsilon_n$ independently of the value of the electron-electron energy $U$. Following the argument used for the two-electron system, we conclude that the quadruplet states with lower spin projections are degenerate with the fully polarized ones. 
%
For completeness, we explicitly write them in terms of second-quantized fermionic operators. They are obtained by acting with the lowering operator $\hat S_-$ on the fully polarized state of the quadruplet, which has the maximal spin projection $S_z = 3/2$,
\begin{equation}
    \ket{Q^{3/2}_{(\ell, m, n)}} =\hat{\varphi}_{\ell \uparrow}^{\dagger} \hat{\varphi}_{m\uparrow}^{\dagger} \hat{\varphi}_{n \uparrow}^{\dagger}\ket{\emptyset}~,
\end{equation}
with $\ell<m<n$. Hence, we calculate
\begin{equation}
    \hat{S}_{-}\ket{Q^{3/2}_{(\ell, m ,n)}}=\sum_p \hat{\varphi}_{p \downarrow}^{\dagger} \hat{\varphi}_{p \uparrow} \hat{\varphi}_{\ell \uparrow}^{\dagger} \hat{\varphi}_{m \uparrow}^{\dagger} \hat{\varphi}_{n \uparrow}^{\dagger}\ket{\emptyset}
\end{equation}
where $p = \ell, m ,n$. For example, taking $p = \ell$ and using the relation $\hat{\varphi}_{p s_z} \hat{\varphi}_{p s_z}^{\dagger} = 1 - \hat{n}_{p s_z}$, we obtain
\begin{equation}
    \hat{\varphi}_{\ell \downarrow}^{\dagger} \hat{\varphi}_{\ell \uparrow} \hat{\varphi}_{\ell \uparrow}^{\dagger} \hat{\varphi}_{m \uparrow}^{\dagger} \hat{\varphi}_{n \uparrow}^{\dagger}\ket{\emptyset}
    = \hat{\varphi}_{\ell \downarrow}^{\dagger}\left(1 - \hat{n}_{\ell \uparrow}\right)\hat{\varphi}_{m \uparrow}^{\dagger} \hat{\varphi}_{n \uparrow}^{\dagger}\ket{\emptyset}
    = \hat{\varphi}_{\ell \downarrow}^{\dagger} \hat{\varphi}_{m \uparrow}^{\dagger} \hat{\varphi}_{n \uparrow}^{\dagger}\ket{\emptyset}~,
\end{equation}
and the same reasoning applies for $p = m$ and $p = n$.
%
Thus, we obtain 
\begin{equation}
    \hat{S}_{-}\ket{Q_{(\ell, m,n)}^{3/2}}=(\hat{\varphi}_{\ell \downarrow}^{\dagger} \hat{\varphi}_{m \uparrow}^{\dagger} \hat{\varphi}_{n \uparrow}^{\dagger}+\hat{\varphi}_{\ell \uparrow}^{\dagger} \hat{\varphi}_{m \downarrow}^{\dagger} \hat{\varphi}_{n \uparrow}^{\dagger}+\hat{\varphi}_{\ell \uparrow}^{\dagger} \hat{\varphi}_{m \uparrow}^{\dagger} \hat{\varphi}_{n \downarrow}^{\dagger})\ket{\emptyset}~,
\end{equation}
and the normalized quadruplet state with $S_z = 1/2$ is given by
\begin{equation}
    \ket{Q_{(\ell, m, n)}^{1/2}}=\frac{1}{\sqrt{3}}(\hat{\varphi}_{\ell \downarrow}^{\dagger} \hat{\varphi}_{m \uparrow}^{\dagger} \hat{\varphi}_{n \uparrow}^{\dagger}+\hat{\varphi}_{\ell \uparrow}^{\dagger} \hat{\varphi}_{m \downarrow}^{\dagger} \hat{\varphi}_{n \uparrow}^{\dagger}+\hat{\varphi}_{\ell \uparrow}^{\dagger} \hat{\varphi}_{m \uparrow}^{\dagger} \hat{\varphi}_{n \downarrow}^{\dagger})\ket{\emptyset}
\end{equation}
Similarly, we can also obtain the normalized quadruplet state with $S_z = -1/2$  as
\begin{equation}
    \left|Q_{(\ell, m,n)}^{-1/2}\right\rangle=\frac{1}{\sqrt{3}}(\hat{\varphi}_{\ell \downarrow}^{\dagger} \hat{\varphi}_{m \downarrow}^{\dagger} \hat{\varphi}_{n \uparrow}^{\dagger}+\hat{\varphi}_{\ell \downarrow}^{\dagger} \hat{\varphi}_{m \uparrow}^{\dagger} \hat{\varphi}_{n \downarrow}^{\dagger}+\hat{\varphi}_{\ell \uparrow}^{\dagger} \hat{\varphi}_{m \downarrow}^{\dagger} \hat{\varphi}_{n \downarrow}^{\dagger})\ket{\emptyset}~,
\end{equation}
and the fully polarized quadruplet state with $S_z = -3/2$ as
\begin{equation}
    \ket{Q_{(\ell, m,n)}^{-3/2}}=\hat{\varphi}_{\ell \downarrow}^{\dagger} \hat{\varphi}_{m \downarrow}^{\dagger} \hat{\varphi}_{n \downarrow}^{\dagger}\ket{\emptyset}~.
\end{equation}
The arguments above are easily generalized to the four-electron case. In this regime, the total spin can be $S = 0, 1, 2$, which corresponds to singlet, triplet, and pentuplet states, respectively. In particular, the full Hilbert space of the system can be decomposed as
\begin{equation}
    \mathbf{H}=   \bigoplus_{i=-2}^{2}\mathbf{P}_{i} \bigoplus_{i=-1}^{1} \mathbf{T}_{i}\bigoplus \mathbf{S}_0~,
\end{equation}
where the dimensions of the subspaces are $\rm{dim}(\mathbf{P}_i) = 1$, $\rm{dim}(\mathbf{T}_i) = 15$,  and $\rm{dim}(\mathbf{S}_0) = 20$, which leads to a total Hilbert space dimension of $\rm{dim}(\mathbf{H}) = 70$.
The pentuplet subspace is characterized by a single, fully polarized state with maximum spin projection $S_z = 2$, which is expressed in terms of fermionic creation operators for the electrons in the eigenstates of $\hat{T}$ with $\ell = 0, m = 1, n = 2, r = 3$ as 
\begin{equation}
    \ket{P_{(\ell, m, n, r)}^2}=\hat{\varphi}_{\ell \uparrow}^{\dagger} \hat{\varphi}_{m \uparrow}^{\dagger} \hat{\varphi}_{n \uparrow}^{\dagger} \hat{\varphi}_{r \uparrow}^{\dagger}\ket{\emptyset}~,
\end{equation}
from which one obtains the states with lower spin projections $S_z =  1, \ldots, -2$. In particular, by applying the operator  $\hat S_-$, 
one obtains the degenerate eigenstate with $S_z = 1$,
\begin{equation}
    \ket{P_{(\ell, m, n, r)}^1} = \frac{1}{2}\bigl(\hat{\varphi}_{\ell \downarrow}^{\dagger} \hat{\varphi}_{m \uparrow}^{\dagger} \hat{\varphi}_{n \uparrow}^{\dagger} \hat{\varphi}_{r \uparrow}^{\dagger}
    + \hat{\varphi}_{\ell \uparrow}^{\dagger} \hat{\varphi}_{m \downarrow}^{\dagger} \hat{\varphi}_{n \uparrow}^{\dagger} \hat{\varphi}_{r \uparrow}^{\dagger}
    + \hat{\varphi}_{\ell \uparrow}^{\dagger} \hat{\varphi}_{m \uparrow}^{\dagger} \hat{\varphi}_{n \downarrow}^{\dagger} \hat{\varphi}_{r \uparrow}^{\dagger}
    + \hat{\varphi}_{\ell \uparrow}^{\dagger} \hat{\varphi}_{m \uparrow}^{\dagger} \hat{\varphi}_{n \uparrow}^{\dagger} \hat{\varphi}_{r \downarrow}^{\dagger}\bigr)\ket{\emptyset}~,
\end{equation}
as well as the degenerate eigenstate with $S_z = 0$,
\begin{equation}
\begin{aligned}
    \ket{P_{(\ell, m, n, r)}^0} =\frac{1}{\sqrt{6}}(&\hat{\varphi}_{\ell \downarrow}^{\dagger} \hat{\varphi}_{m \downarrow}^{\dagger} \hat{\varphi}_{n \uparrow}^{\dagger} \hat{\varphi}_{r \uparrow}^{\dagger}+\hat{\varphi}_{\ell \downarrow}^{\dagger} \hat{\varphi}_{m \uparrow}^{\dagger} \hat{\varphi}_{n \downarrow}^{\dagger} \hat{\varphi}_{r \uparrow}^{\dagger}+\hat{\varphi}_{\ell \downarrow}^{\dagger} \hat{\varphi}_{m \uparrow}^{\dagger} \hat{\varphi}_{n \uparrow}^{\dagger} \hat{\varphi}_{r \downarrow}^{\dagger} +\\
    &+\hat{\varphi}_{\ell \uparrow}^{\dagger} \hat{\varphi}_{m \downarrow}^{\dagger} \hat{\varphi}_{n \downarrow}^{\dagger} \hat{\varphi}_{r \uparrow}^{\dagger}+\hat{\varphi}_{\ell \uparrow}^{\dagger} \hat{\varphi}_{m \downarrow}^{\dagger} \hat{\varphi}_{n \uparrow}^{\dagger} \hat{\varphi}_{r \downarrow}^{\dagger}+\hat{\varphi}_{\ell \uparrow}^{\dagger} \hat{\varphi}_{m \uparrow}^{\dagger} \hat{\varphi}_{n \downarrow}^{\dagger} \hat{\varphi}_{r \downarrow}^{\dagger})\ket{\emptyset}~.
\end{aligned}
\end{equation}
The corresponding degenerate states with $S_z = -2$ and $S_z = -1$ can be generated from those with $S_z = 2$ and $S_z = 1$ by flipping each individual spin $z$-projection ($\{\uparrow,\downarrow\} \to \{\downarrow,\uparrow\}$).
Using the argument based on the Pauli Principle given above, we find that within the pentuplet subspace the eigenenergies are independent of the electron–electron interaction energy $U$.

\section{Van Vleck paramagnetism}\label{app:van_vleck_paramagnetism}
In this Appendix, we briefly recall for the reader the microscopic theory underlying Van Vleck paramagnetism~\cite{Grosso_2014}.
We consider a system consisting of $N_e$ interacting electrons described by the non-relativistic Hamiltonian
\begin{equation}\label{eq:vv-H0}
  \hat H_0=\sum_{i=1}^{N_e}\frac{\hat{\bm p}_i^{2}}{2m} + \sum_{i=1}^{N_e}V(\hat{\bm r}_i)
  + \frac{1}{2} \sum_{i\neq j} \frac{e^{2}}{|\hat{\bm r}_i-\hat{\bm r}_j|}\, ,
\end{equation}
where $V(\hat{\bm r}_i)$ is the electron-nuclear potential and the last term is the electron-electron Coulomb repulsion; spin-orbit coupling is neglected. We introduce a uniform magnetic field $\bm B$ in the symmetric gauge, $\bm A(\bm r)=\frac{1}{2} \bm B\times\bm r$, through the minimal substitution $\hat{\bm p}_i\to\bm \hat{\bm p}_i+(e/c)\bm A(\hat{\bm r}_i)$ in the kinetic energy, supplemented by the Zeeman coupling $2\mu_{\rm B}{\bm B}\cdot\hat{\bm s}_i$ of each electronic spin, where $\mu_{\rm B}=e /(2 m c)$ denotes the Bohr magneton (with $\hbar=1$). By defining the total orbital and spin angular momenta
\begin{subequations}
\begin{align}
  \hat{\bm L}&=\sum_{i=1}^{N_e}\hat{\bm r}_i \times \hat{ \bm p}_i~, \\
  \hat{\bm S}&=\sum_{i=1}^{N_e}\hat{\bm s}_i~, 
\end{align}
\end{subequations}
one obtains
\begin{equation}
   \hat H=\hat H_0 +\mu_{\rm B} \bm B\cdot(\hat{\bm L}+2\hat{\bm S}) + \frac{e^{2}}{8mc^{2}}\sum_{i=1}^{N_e}\big(\bm B \times \hat{\bm r}_i\big)^{2} ~.
\end{equation}
The first correction represents the Zeeman operator $-\hat{\bm \mu} \cdot \bm B$, with magnetic dipole $\hat{\bm \mu}=-\mu_{\rm B}(\hat{\bm L} + 2\hat{\bm S})$; the second is diamagnetic and quadratic in $\bm B$. Both terms are assumed to be small compared with $\hat H_0$ and are treated by perturbation theory. Moreover, we consider a system whose ground state $|\Psi_0\rangle$ is non-degenerate. The first-order Zeeman shift vanishes, since $\langle\Psi_0|\hat{\bm L}|\Psi_0\rangle = \langle\Psi_0|\hat{\bm S}|\Psi_0\rangle=0$ in a TRS ground state, signaling the absence of a permanent magnetic moment. Collecting all contributions up to the second-order in $\bm B$, one obtains
\begin{equation}
    \Delta E_0 = \Delta E_{\rm D}+\Delta E_{\rm P}
\end{equation}
with the (positive) diamagnetic term
\begin{equation}
  \Delta E_{\rm D} = \frac{e^{2}}{8mc^{2}}
  \Bigl\langle\Psi_0 \Bigl|\sum_{i=1}^{N_e} \bigl(\bm B\times\hat{\bm r}_i\big)^{2} \Bigr|\Psi_0\Bigr\rangle > 0~,
\end{equation}
whereas the (negative) second-order term is called the Van Vleck contribution and is given by 
\begin{equation}\label{eq:param_susc}
  \Delta E_{\rm P}= -\mu_{\rm B}^2 \sum_{n \neq0} \frac{|\langle\Psi_n|\bm B\cdot(\hat{\bm L}+2\hat{\bm S})|\Psi_0\rangle|^{2}}{E_n-E_0}<0~.
\end{equation}
Accordingly, in the regime where the ground state and first excited state are nearly degenerate and are coupled via the paramagnetic magnetization operator, the paramagnetic contribution to the magnetic response can dominate the diamagnetic contribution, thereby driving the system into a paramagnetic regime.
This mechanism, referred to as the Van Vleck mechanism, underlies the increased paramagnetic behavior of individual molecules and drives the photon-condensation phase transition in our system.
In our system, we find that states belonging to subspaces with different total spin squared $S^2$ are not coupled by the orbital paramagnetic magnetization operator. Consequently, the corresponding matrix element in Eq.~\eqref{eq:param_susc} vanishes, and such states cannot contribute to the Van Vleck mechanism. Accidental degeneracies between states with different $S^2$ therefore do not enhance the orbital paramagnetic response.

\section{Bosonization and Polaritons}\label{app:bosonization}

To derive the polaritonic spectrum of the system, we start from the full Hamiltonian expressed in Eq.~\eqref{eq:Hfull} and replace $\hat{a} \rightarrow \bar{\alpha} \sqrt N+\delta \hat a$ and $\hat{a}^\dagger \rightarrow \bar{\alpha} \sqrt N+\delta \hat a^\dagger$, where $\bar{\alpha}$ is solution of Eq.~\eqref{eq:selfcons}, and $\delta \hat{a}$ and $\delta \hat{a}^\dagger$ are a zero-average fluctuations on top of the mean-field solution. 
The shifted Hamiltonian reads
\begin{equation}
\begin{aligned}
    \hat{\mathcal H}_{\rm S}&=\omega_{\rm c}(\sqrt{N} \bar{\alpha}+\delta \hat{a}^\dagger)(\sqrt{N} \bar{\alpha}+\delta \hat{a})-\sum_{k, j, s_z} t_{j, j+1} e^{-i \eta(2 \sqrt{N} \bar{\alpha}+\delta \hat{a}+\delta \hat{a}^\dagger) / \sqrt{N}} \hat c_{j+1 s_z k}^\dagger \hat c_{j s_z k} +\text{ H.c.} + \hat{U}_k\\ \nonumber
    &= \omega_{\rm c}\left[N \bar{\alpha}^2+\sqrt{N} \bar{\alpha}(\delta \hat a+\delta\hat a^\dagger)+\delta \hat a^{\dagger} \delta \hat a\right]-\sum_{k, j, s_z} t_{j, j+1} e^{-i \eta(2 \sqrt{N} \bar{\alpha}+\delta \hat{a}+\delta \hat{a}^\dagger) / \sqrt{N}} \hat c_{j+1 s_z k}^\dagger \hat c_{j s_z k} +\text{ H.c.} + \hat{U}_k~.
\end{aligned}
\end{equation}
Since we are looking at small fluctuations over the mean-field ground state, we expand the Hamiltonian up to second-order in the $\delta \hat a$ and $\delta \hat a^\dagger$ terms, obtaining 
\begin{align}\label{shifted_H}
	\hat{\mathcal {H}_{\rm S}}&=  \omega_{\rm c}\left[N \bar{\alpha}^2+\sqrt{N} \bar{\alpha}(\delta \hat a+\delta\hat a^\dagger)+\delta \hat a^{\dagger} \delta \hat a\right] + \sum_{k = 1}^{N} \hat{H}_{{\rm{MF}}, k }(\bar{\alpha})\nonumber \\
	&+ \frac{\sqrt{N}}{2} \hat{\mathcal M}_{\rm p}(\bar{\alpha})(\delta\hat a + \delta \hat a ^\dagger) + \frac{1}{8}\hat{\mathcal M}_{\rm d}(\bar{\alpha}) (\delta \hat a + \delta \hat{a}^\dagger)^2~,
\end{align}
where $\hat{\mathcal M}_{\rm p}$  ($\hat{\mathcal M}_{\rm d}$) is the paramagnetic (diamagnetic) component of the total magnetization, defined in the main text.  
The first term in \eqref{shifted_H} involves the electromagnetic degrees of freedom solely.
The second term is the mean-field Hamiltonian $\hat{H}_{{\rm{MF}}, k }(\bar{\alpha})$, written in its eigenbasis as
\begin{equation}
\sum_{k=1}^{N} \hat{H}_{{\rm{MF}}, k}(\bar{\alpha}) = \sum_{k = 1}^N\sum_{\ell} E_\ell(\bar{\alpha})  |\Psi_{\ell k}(\bar{\alpha})\rangle \langle \Psi_{lk}(\bar{\alpha})|~,
\end{equation}
%
%
%
Using the resolution of the identity $\mathbb{1} = \sum_{\ell}|\Psi_{\ell k}(\bar{\alpha})\rangle \langle \Psi_{\ell k}(\bar{\alpha})|$, the third term in Eq.~\eqref{shifted_H} can be rewritten as
\begin{equation}
	\frac{1}{2\sqrt{N}} (\delta\hat a + \delta \hat a ^\dagger) \sum_{k = 1}^{N} \sum_{\ell,m} \mathcal M^{\ell,m}_{\rm p}(\bar{\alpha}) |\Psi_{\ell k}(\bar{\alpha})\rangle \langle \Psi_{mk}(\bar{\alpha})|~,
\end{equation}
where we defined
\begin{equation}
	\mathcal M^{\ell,m}_{\rm p}(\bar{\alpha}) \equiv \sum_{k=1}^N\langle\Psi_{\ell k}(\bar{\alpha})| \hat{\mathcal {M}}_{p}(\bar{\alpha})|\Psi_{mk}(\bar{\alpha})\rangle~.
\end{equation}
Because all molecules are identical, the matrix elements above do not depend on $k$, and the summation therefore produces an overall factor of $N$.
Similarly,  the last term in Eq.~\eqref{shifted_H} can be written as
\begin{equation}
	\frac{1}{8N}(\delta \hat a+\delta \hat a^{\dagger})^2 \sum_{k = 1}^{N} \sum_{\ell, m} \mathcal{M}_{\rm{d}}^{\ell, m}(\bar{\alpha})|\Psi_{\ell k}(\bar{\alpha})\rangle \langle \Psi_{mk}(\bar{\alpha})|~.
\end{equation}
Hence, explicitly, one has 
\begin{align*}
	\hat{\mathcal{H}_{\rm S}}&=  \omega_{\rm c}\left[N \bar{\alpha}^2+\sqrt{N} \bar{\alpha}(\delta \hat a+\delta\hat a^\dagger)+\delta \hat a^{\dagger} \delta \hat a\right] 
	+ \sum_{k = 1}^N\sum_{\ell} E_\ell(\bar{\alpha})  |\Psi_{\ell k}(\bar{\alpha})\rangle \langle \Psi_{\ell k}(\bar{\alpha})| \\
	&+ \frac{1}{2\sqrt{N}} (\delta\hat a + \delta \hat a ^\dagger) \sum_{k=1}^{N} \sum_{\ell,m} \mathcal M^{\ell ,m}_{\rm p}(\bar{\alpha}) |\Psi_{\ell k}(\bar{\alpha})\rangle \langle \Psi_{mk}(\bar{\alpha})|+ \frac{1}{8N}(\delta \hat a+\delta \hat a^{\dagger})^2 \sum_{k = 1}^{N} \sum_{\ell, m} \mathcal{M}_{\rm{d}}^{\ell, m}(\bar{\alpha})|\Psi_{\ell k}(\bar{\alpha})\rangle \langle \Psi_{mk}(\bar{\alpha})|~.
\end{align*}
By using the collective notation
\begin{equation}\label{eq:Sigmaoperators}
    \hat{\Sigma}_{\ell, m} \equiv \sum_k |\Psi_{\ell k}(\bar{\alpha})\rangle \langle \Psi_{mk}(\bar{\alpha})|~,
\end{equation}
the Hamiltonian assumes the more compact form
\begin{align}
	\hat{\mathcal{H}_{\rm S}}&=  \omega_{\rm c}\left[N \bar{\alpha}^2+\sqrt{N} \bar{\alpha}(\delta \hat a+\delta\hat a^\dagger)+\delta \hat a^{\dagger} \delta\hat a\right] + \sum_{\ell }^{} E_\ell(\bar{\alpha})\hat{\Sigma}_{\ell , \ell}  \nonumber \\
	&+ \frac{1}{2\sqrt{N}} (\delta\hat a + \delta \hat a ^\dagger) \sum_{\ell,m} \mathcal M^{\ell,m}_{\rm p}(\bar{\alpha})\hat{\Sigma}_{\ell, m}+ \frac{1}{8N}(\delta \hat a+\delta \hat a^{\dagger})^2  \sum_{\ell, m} \mathcal{M}_{\rm d}^{\ell, m}(\bar{\alpha}) \hat{\Sigma}_{\ell, m}~.
\end{align}
Proceeding as in Refs.~\cite{kurucz_pra_2010, Mercurio_PRR_2024}, we focus on the symmetric subspace of the Hilbert space of the $N$ molecules,
%
which is spanned by  the occupation number states defined as
%
\begin{equation}
\begin{aligned}
    |n_0, m_1, \ldots,p_{s-1}\rangle = \frac{1}{\sqrt{n!m! \cdots p!}} \sum_{\rm{perm}}&|\Psi_{0\,1}(\bar{\alpha})\rangle\cdots |\Psi_{0\,n_0}(\bar{\alpha})\rangle \cdots |\Psi_{1\, n_0 + 1}(\bar{\alpha})\rangle\cdots |\Psi_{1\, n_0 + m_1}(\bar{\alpha})\rangle \otimes \\
    &\otimes|\Psi_{s-1\,N-p+1}(\bar{\alpha})\rangle \cdots  |\Psi_{s-1\,N}(\bar{\alpha})\rangle~,
\end{aligned}
\end{equation}
where each basis state is a symmetric linear combination of all possible electronic configurations of $N$ molecules in which there are  $n$ molecules in the many-body state $|\Psi_{0\, k}(\bar{\alpha})\rangle$, $m$ molecules in the many-body state $|\Psi_{1\, k}(\bar{\alpha})\rangle$, $\ldots$, and $p$ molecules in the many-body state $|\Psi_{s-1\, k}(\bar{\alpha})\rangle$, where $s$ denotes the number of many-body eigenstates of a single molecule. The non-negative indices $n_0, m_1,\ldots,p_{s-1}$  satisfy the constraint $n_0 + m_1+ \ldots p_{s-1} = N$.
By applying an occupation number state on the collective operators defined in Eq.~\eqref{eq:Sigmaoperators}, we obtain the following properties
\begin{subequations}
\begin{align}
\hat{\Sigma}_{\ell,\ell'}\ket{n_\ell,m_{\ell'},\ldots}&=\sqrt{(n+1)m} 
\ket{(n+1)_\ell,m_{\ell'}-1,\ldots}, \label{eq:Sigma_prop1} \\
\hat{\Sigma}_{\ell,\ell}\ket{n_\ell,m_{\ell'},\ldots}&=n\ket{n_\ell,m_{\ell'},\ldots}, \label{eq:Sigma_prop2}
\end{align}
\end{subequations}
where $\ell \neq \ell'$. 
Moreover, the collective operators satisfy the commutation relation $[\hat{\Sigma}_{\ell,\ell'}, \hat{\Sigma}_{m,m'}]=\delta_{\ell',m}\hat{\Sigma}_{\ell,m'}-
\delta_{m',\ell}\hat{\Sigma}_{m,\ell'}$.
Here, we define $s-1$ pairs of bosonic creation and annihilation operators, $\hat{b}_\ell^\dagger$ and $\hat{b}_\ell$ such that
\begin{subequations}
\begin{align}
\hat{b}_\ell \ket{n_0,m_\ell,\cdots}&=\sqrt{m}
 \ket{(n+1)_0,(m-1)_\ell,\cdots}~, \label{eq:b_prop1}\\
 \hat{b}_\ell^\dagger \hat{b}_\ell \ket{n_0,m_\ell,\cdots}&=m
\ket{n_0,m_\ell,\cdots}~,  \label{eq:b_prop2}
\end{align}
\end{subequations}
where $\ell=1,\ldots,s-1$, and the mean-field ground state $\ket{F_0(\bar{\alpha})}=\ket{N_0,0,\cdots,0}=\prod^N_{k=1} \ket{\Psi_{0 k}(\bar{\alpha})}$, which describes all molecules occupying the state $\ket{\Psi_0(\bar{\alpha})}$, acts as the vacuum state with respect to every operator $\hat{b}_\ell$. 
The creation operators $\hat{b}^\dagger_l$ applied on the vacuum state $\ket{F_0(\bar{\alpha})}$ describe the collective matter excitations, which represent the bright modes.
By comparing Eqs.~\eqref{eq:Sigma_prop1}-\eqref{eq:Sigma_prop2} with Eqs.~\eqref{eq:b_prop1}-\eqref{eq:b_prop2}, we write the collective operators
 in terms of the bosonic fields according to a multilevel Holstein-Primakoff transformation~\cite{kurucz_pra_2010}, 
\begin{subequations}
\begin{align}
\hat{\Sigma}_{0,0}&=N-\sum_{\ell>0} \hat{b}_\ell^\dagger\hat{b}_\ell~,\\
\hat{\Sigma}_{\ell,0}&=\hat{b}_\ell^\dagger \sqrt{N-\sum_{\ell^\prime>0} \hat{b}_{\ell^\prime}^\dagger\hat{b}_{\ell^\prime} }~,\\
\hat{\Sigma}_{\ell,\ell^\prime}&=\hat{b}_\ell^\dagger\hat{b}_{\ell^\prime}~,
\end{align}
\end{subequations}
where $\ell, \ell^\prime>0$.
%
%


%
%

%
In the proximity of the mean-field matter ground state $\ket{F_0(\bar{\alpha})}$, that is, for a small number of collective matter excitations, one may use the approximation $\hat\Sigma_{\ell,0} \simeq \sqrt{N}\hat b_\ell^\dagger$ and discard $\hat\Sigma_{\ell,\ell'}$.
This results in a quadratic Hamiltonian in the bosonic fields, describing light and matter excitations,
\begin{equation}\label{eq:Hquadratic}
\begin{aligned}
\hat{\mathcal{H}_{\rm S}} &\simeq  \omega_{\rm c} \delta \hat{a}^{\dagger} \delta \hat{a}+ \frac{\sqrt{N}}{2}\left[2 \omega_{\rm c}\bar{\alpha} + \mathcal{M}_{\rm p}^{0, 0}(\bar{\alpha})\right](\delta \hat{a}+\delta \hat{a}^{\dagger}) \\
& +N \omega_{\rm c} \bar{\alpha}^2+N E_0 (\bar{\alpha})+\sum_{\ell > 0} [E_\ell(\bar{\alpha})-E_0(\bar{\alpha}) ] \hat{b}_\ell^{\dagger} \hat{b}_\ell \\
& +\frac{1}{2}(\delta \hat{a}+\delta \hat{a}^{\dagger}) \sum_{\ell > 0}[\mathcal{M}_{\rm p}^{\ell, 0}(\bar{\alpha}) \hat{b}_\ell^{\dagger}+\mathcal{M}_{\rm p}^{0, \ell}(\bar{\alpha}) \hat{b}_\ell]  +\frac{1}{8}(\delta \hat{a}+\delta \hat{a}^{\dagger})^2 \mathcal{M}_{\rm d}^{0,0}(\bar{\alpha})~.
\end{aligned}
\end{equation}
We notice that the collective operators $\Sigma_{\ell,m}$, and equivalently the bright mode operator $\hat{b}_{\ell}$ and $\hat{b}_{\ell}^\dagger$, leave both the $z$ component of the spin and the total spin $S$ unchanged. As a result, the polaritonic excitations can be analyzed separately within each sector characterized by the total spin quantum numbers $S$ and $S_z$.
The square bracket in the first line of Eq.~\eqref{eq:Hquadratic} vanishes because $\bar{\alpha}$ satisfies the nonlinear Eq.~\eqref{eq:selfcons}. To make this explicit, one can use the Hellman-Feynman theorem, obtaining
%
%
\begin{equation}
\begin{aligned}
\partial_\alpha E_0(\alpha)  & =\frac{1}{N} \sum_{k=1}^N \langle F_0(\alpha)| \partial_\alpha\hat{H}_{{\rm{MF}}, k}(\alpha)|F_0(\alpha)\rangle =\langle F_0(\alpha)| \hat{\mathcal{M}}_{\rm p}(\alpha)|F_0(\alpha)\rangle=\mathcal{M}_{\rm p}^{0,0} (\alpha)~,
\end{aligned}
\end{equation}
 $\bar \alpha$ is equivalently the solution of the nonlinear equation $ 2 \omega_{\rm c} \bar\alpha +\mathcal{M}_{\rm p}^{0,0} (\bar\alpha) = 0$.

Thus, by expanding the shifted Hamiltonian around $\alpha=\bar{\alpha}$, one obtains the quadratic polaritonic Hamiltonian, which can be written as
\begin{equation}
\begin{aligned}
	\hat{\mathcal{H}}_{\rm{pol}} & =\omega_{\rm c} \delta \hat a^\dagger \delta \hat{a}+\sum_{\ell > 0}[E_\ell(\bar \alpha)-E_0(\bar \alpha)] \hat b_\ell^{\dagger} \hat{b}_\ell  +\frac{1}{2}(\delta \hat{a}+\delta \hat a^\dagger) \sum_{\ell > 0}[\mathcal{M}_{\rm{p}}^{\ell, 0}(\bar \alpha) \hat{b}_\ell^{\dagger}+\mathcal{M}_{\rm{p}}^{0, \ell}(\bar\alpha) \hat{b}_\ell]  +\frac{1}{8} (\delta\hat{a}+\delta \hat a^\dagger)^2 \mathcal{M}_{\rm d}^{0, 0}(\bar \alpha)~.
\end{aligned}
\end{equation}
Polaritons are hybrid excitations of light and matter degrees of freedom and are described by the following linear combination
\begin{equation}
	\hat{p}_\nu=X_\nu \delta \hat{a}+Y_\nu \delta \hat{a}^{\dagger}+\sum_{\ell > 0}\left(W_{\nu, \ell} \hat{b}_\ell+Z_{\nu, \ell} \hat{b}_\ell^{\dagger}\right)~.
\end{equation}
Since the polariton represents a proper bosonic excitation of the system, the operator $\hat p_\nu$ is a bosonic operator that obeys the equation of motion $[\hat{\mathcal{H}}_{\rm{pol}}, \hat{p}_\nu] = -\Omega_{\rm{p}, \nu}\,\hat{p}_\nu$.
Since $\hat p_\nu$ is expressed as a linear combination of all bosonic creation and annihilation operators, we evaluate the commutator of each of these operators with the polaritonic Hamiltonian.
Thus, for the photon operators, one has
\begin{equation}
    \begin{aligned}
    [\hat{\mathcal{H}}_{\rm{pol}}, \delta\hat {a}] &= -\omega_{\rm c} \delta \hat{a}-\frac{1}{4} \mathcal{M}_{\rm d}^{0,0}(\bar\alpha)(\delta \hat{a}+\delta \hat{a}^{\dagger}) -\frac{1}{2} \Big(\sum_{\ell > 0}\mathcal{M}_{\rm p}^{\ell,0}(\bar\alpha) \hat{b}_\ell^{\dagger}+\mathcal{M}_{\rm p}^{0, \ell}(\bar\alpha) \hat{b}_\ell\Big)\\    
    [\hat{\mathcal{H}}_{\rm{pol}}, \delta \hat{a}^{\dagger}] &=  \omega_{\rm c} \delta \hat{a}^{\dagger}+\frac{1}{4} \mathcal{M}_{\rm d}^{0,0}(\bar\alpha)(\delta \hat{a}+\delta \hat a^{\dagger}) +\frac{1}{2} \sum_{\ell > 0}\Big(\mathcal{M}_{\rm p}^{\ell, 0}(\bar\alpha) \hat{b}_\ell^{\dagger}+\mathcal{M}_{\rm p}^{0, \ell}(\bar\alpha) \hat{b}_\ell\Big)~,\\
    \end{aligned}
\end{equation}
%
%
while for the bright modes operators, one finds
\begin{equation}
    \begin{aligned}
    [\hat{\mathcal{H}}_{\rm{pol}}, \hat{b}_\ell]&= -[E_\ell(\bar\alpha)-E_0(\bar\alpha)] \hat{b}_\ell-\frac{1}{2} \mathcal{M}_{\rm {p}}^{\ell, 0}(\bar\alpha)(\delta \hat{a}+\delta \hat{a}^{\dagger})~,\\
        [\hat{\mathcal{H}}_{\rm{pol}}, \hat{b}_\ell^{\dagger}]&=[E_\ell(\bar\alpha)-E_0(\bar\alpha)] \hat{b}_\ell^{\dagger}+\frac{1}{2} \mathcal{M}_{\rm p}^{0, \ell}(\bar\alpha)(\delta \hat{a}+\delta \hat{a}^{\dagger})~.
    \end{aligned}
\end{equation}
%
%
%

The equation of motion can be mapped into the eigenvalues problem $\Xi \bm v_\nu= \Omega_{\rm{p}, \nu} \bm v_\nu$, being $\bm v_\nu= (X_\nu, Y_\nu,\bm W_\nu, \bm Z_\nu)^{T}$. 
This is achieved by carrying out the following
\begin{equation}
\begin{aligned}
    [\hat{\mathcal{H}}_{\rm{pol}}, \hat{p}_\nu]&=  X_\nu [\hat{\mathcal{H}}_{\rm{pol}},\delta \hat{a}]+Y_\nu [\hat{\mathcal{H}}_{\rm{pol}},\delta \hat{a}^{\dagger}]+\sum_{\ell > 0}\Big(W_{\nu, \ell} [\hat{\mathcal{H}}_{\rm{pol}},\hat{b}_\ell]+Z_{\nu, \ell} [\hat{\mathcal{H}}_{\rm{pol}},\hat{b}_\ell^{\dagger}\Big)] \\
    &=\Omega_{\rm{p}, \nu} \left[ X_\nu \delta \hat{a}+Y_\nu \delta \hat{a}^{\dagger}+\sum_{\ell > 0}\Big(W_{\nu, \ell} \hat{b}_\ell+Z_{\nu, \ell} \hat{b}_\ell^{\dagger}\Big)\right] ~.
\end{aligned}    
\end{equation}
%
The resulting $2s \times 2s$ matrix $\Xi$, written in the basis $(\delta \hat{a}, \delta\hat{a}^\dagger, \hat{b}_\ell, \hat{b}_\ell^\dagger)$, is explicitly expressed as
\begin{equation}\label{hopfield}
\Xi=\left(\begin{array}{cccc}
	\omega_{\rm c}+\dfrac{1}{4}\mathcal M_{\rm d}^{0,0}(\bar\alpha) & - \dfrac{1}{4}\mathcal M_{\rm d}^{0,0} (\bar\alpha)& \bm g^*(\bar\alpha) & -\bm g(\bar\alpha) \\[1.75ex]
	\dfrac{1}{4}\mathcal M_{\rm d}^{0,0}(\bar\alpha) & -\omega_{\rm c}-\dfrac{1}{4}\mathcal M_{\rm d}^{0,0}(\bar\alpha) & \bm g^*(\bar\alpha) & -\bm g(\bar\alpha) \\[1.75ex]
	\bm g^{*T}(\bar\alpha) & -\bm g^{*T}(\bar\alpha) & \bm{\Omega}(\bar\alpha) & 0 \\[1.75ex]
	\bm g^{T}(\bar\alpha) & -\bm g^{T}(\bar\alpha) & 0 & -\bm{\Omega}(\bar\alpha)
\end{array}\right)~,
\end{equation}
where 
\begin{align*}
    \bm g &= \dfrac{1}{2}[\mathcal{M}_{\rm p}^{1,0}(\bar\alpha), \mathcal{M}_{\rm p}^{2,0}(\bar\alpha), \ldots, \mathcal{M}_{\rm p}^{s -1 ,0}(\bar\alpha)]~,\\
    \bm \Omega &= {\rm diag}[E_1(\bar \alpha) - E_0(\bar\alpha), E_2(\bar \alpha) - E_0(\bar\alpha), \ldots, E_{s-1}(\bar\alpha)- E_0(\bar\alpha)]~.
\end{align*}
The Hopfield matrix $\Xi$ in Eq. \eqref{hopfield} has been diagonalized numerically, and its eigenvalues correspond to the polaritonic excitation energies.

\end{widetext}
\bibliography{biblio}%
\end{document}